%% file: S2RDF_TR.tex
\documentclass{sig-alternate}
\pdfoutput=1

\usepackage{balance}  
\usepackage{float}
\usepackage{comment}
\usepackage{graphicx}
\usepackage{amssymb}
\usepackage{amsmath}
\usepackage{tabularx}
\usepackage{rotating}
\usepackage{multirow}
\usepackage[lined,ruled,linesnumbered]{algorithm2e}
\usepackage{booktabs}
\usepackage{pgfplots}
\pgfplotsset{compat=newest}
\usetikzlibrary{patterns}
\usepackage{url}
\usepackage{tikz}

\makeatletter
\let\@copyrightspace\relax
\makeatother

\begin{document}


\title{S2RDF: RDF Querying with SPARQL on Spark}

\numberofauthors{4} 
\author{
	\alignauthor
		Alexander Sch\"{a}tzle\\
		\affaddr{University of Freiburg, Germany}\\
		\email{schaetzle@informatik.uni-freiburg.de}
	\alignauthor
		Martin Przyjaciel-Zablocki\\
		\affaddr{University of Freiburg, Germany}\\
		\email{zablocki@informatik.uni-freiburg.de}
	\alignauthor
		Simon Skilevic\\
		\affaddr{University of Freiburg, Germany}\\
		\email{skilevis@informatik.uni-freiburg.de}
	\and  
	\alignauthor
		Georg Lausen\\
		\affaddr{University of Freiburg, Germany}\\
		\email{lausen@informatik.uni-freiburg.de}
}

\maketitle

\begin{abstract}
\input{sections/abstract.tex}
\end{abstract}

\section{Introduction}
\label{sec:Introduction}
\input{sections/introduction.tex}

\section{Foundations}
\label{sec:Foundations}
\input{sections/foundations.tex}

\section{Related Work}
\label{sec:RelatedWork}
\input{sections/related_work.tex}

\section{Relational Mappings for RDF}
\label{sec:RDFMapping}
\input{sections/rdf_mapping.tex}

\section{Extended Vertical Partitioning}
\label{sec:ExtVP}
\input{sections/extvp.tex}

\section{S2RDF Query Processing}
\label{sec:Querying}
\input{sections/querying.tex}

\section{Evaluation}
\label{sec:Evaluation}
\input{sections/evaluation.tex}

\section{Conclusion}
\label{sec:Conclusion}
\input{sections/conclusion.tex}

\bibliographystyle{abbrv}
\bibliography{S2RDF_TR}

\newpage
\mbox{}
\newpage
\appendix
\section{WatDiv Basic Testing}
\label{sec:AppendixWatDivBasic}
\input{sections/appendix_watdiv_basic.tex}

\newpage
\section{WatDiv Selectivity Testing (ST)}
\label{sec:AppendixWatDivSelectivity}
\input{sections/appendix_watdiv_selectivity.tex}

\newpage
\section{WatDiv Incremental Linear (IL)}
\label{sec:AppendixWatDivIncremental}
\input{sections/appendix_watdiv_incremental.tex}

\end{document}

%% file: sections/abstract.tex
RDF has become very popular for semantic data publishing due to its flexible and universal graph-like data model. Yet, the ever-increasing size of RDF data collections makes it more and more infeasible to store and process them on a single machine, raising the need for distributed approaches.
Instead of building a standalone but closed distributed RDF store, we endorse the usage of existing infrastructures for Big Data processing, e.g.~Hadoop.
However, SPARQL query performance is a major challenge as these platforms are not designed for RDF processing from ground. Thus, existing Hadoop-based approaches often favor certain query pattern shape while performance drops significantly for other shapes.
In this paper, we describe a novel relational partitioning schema for RDF data called ExtVP that uses a semi-join based preprocessing, akin to the concept of Join Indices in relational databases, to efficiently minimize query input size regardless of its pattern shape and diameter. Our prototype system S2RDF is built on top of Spark and uses its relational interface to execute SPARQL queries over ExtVP.
We demonstrate its superior performance in comparison to state of the art SPARQL-on-Hadoop approaches using the recent WatDiv test suite. S2RDF achieves sub-second runtimes for majority of queries on a billion triples RDF graph.

%% file: sections/introduction.tex
In recent years, driven by an increasing public interest in \textit{Open Data} publishing and \textit{Semantic Web} initiatives like \textit{Schema.org}, the amount of semantically annotated data has grown steadily towards massive scale with RDF~\cite{rdfprimer} being the W3C standard.
RDF has a very flexible data model consisting of so-called \textit{triples} $(s,p,o)$ that can be interpreted as a labeled directed edge, $s \xrightarrow{p} o$, with $s$ and $o$ being arbitrary resources.
Thus a set of RDF triples forms an interlinkable graph whose flexibility allows to represent a large variety from highly to loosely structured datasets~\cite{duan_apples_2011}.

Nowadays, RDF collections with billions of triples are not unusual, e.g.~\textit{Google Knowledge Vault}~\cite{google_knowledge_vault_2014}, raising the need for distributed solutions.
However, RDF graphs may have very diverse structural properties which makes efficient distributed storing and querying a non-trivial task.
One possible approach is to build a standalone distributed RDF store that is designed from ground for RDF management and comes with its own boxed data store, e.g.~Virtuoso Cluster~\cite{boncz_VirtuosoCluster_2014} or YARS2~\cite{harth_yars2_2007}.
But this also means that data stored in these systems can only be accessed via application specific interfaces or query endpoints which hampers interoperability with other systems and causes high integration costs.

On the other side, there already exist platforms for distributed \textit{Big Data} processing which are also offered on a rental basis by leading \textit{Cloud} providers, e.g.~\textit{Amazon EC2}.
\textit{Hadoop}\footnote{\url{https://hadoop.apache.org/}} has become one of the de facto industry standards in this area.
The key concept of these general-purpose platforms is to have a unified pool for data storage (HDFS for Hadoop) that is shared among various applications on top.
Thus, different systems can access the same data without duplication or movement for various purposes (e.g.~querying, data mining or machine learning).
In our view, to build a system for semantic data processing on top of these existing infrastructures is superior to a specialized deployment in terms of cost-benefit ratio and can provide more synergy benefits.
However, as they are designed for general-purpose data processing and not specifically for RDF data management, the main challenge is to achieve a performance that is in the same order of magnitude compared to specialized systems built from ground for RDF processing.

There exists a large body of work on RDF querying using \textit{MapReduce} as the execution layer, e.g.~\cite{husain_HadoopRDF_2011, rohloff_SHARD_2011, pigsparql_2013, zhang_EAGRE_2013}.
These systems typically have a very good scalability but cannot provide interactive query runtimes due to the batch-oriented nature of MapReduce.
More recently, the emergence of NoSQL key-value stores (e.g.~\textit{HBase}, \textit{Accumulo}) and in-memory frameworks (e.g.~\textit{Impala}, \textit{Spark}) for Hadoop facilitates the development of new systems for RDF querying that are also applicable for more interactive workloads, e.g.~\cite{papailiou_H2RDF+_2013, punnoose_Rya_2012, sempala_2014}.
Yet still, these systems are typically optimized for query patterns with a small diameter like star-shapes and small chains. The performance often drops significantly for unselective queries that produce many (intermediate) results or queries that contain long linear chains.

In this paper, we introduce S2RDF (\underline{S}PARQL on \underline{S}park for \underline{RDF}), a SPARQL query processor based on the in-memory cluster computing framework \textit{Spark}.
It uses the relational interface of Spark for query execution and comes with a novel partitioning schema for RDF called \textit{ExtVP} (\underline{Ext}ended \underline{V}ertical \underline{P}artitioning) that is an extension of the well-known \textit{Vertical Partitioning} (VP) schema introduced by Abadi et al.~\cite{abadi_vp_2007}. ExtVP enables to exclude unnecessary data from query processing by taking into account the possible join correlations between tables in VP.
It is based on \textit{semi-join reductions}~\cite{bernstein_semijoin_1981} and is conceptually related to \textit{Join Indices}~\cite{valduriez_JoinIndices_1987} in relational databases.
In contrast to existing layouts, the optimizations of ExtVP are applicable for all query shapes regardless of its diameter.

Our major contributions are as follows:
(1) We define a novel relational partitioning schema for RDF data called ExtVP that can significantly reduce the input size of a query.
(2) As an optional storage optimization, ExtVP allows to define a selectivity threshold to effectively reduce the size overhead compared to VP while preserving most of its performance benefit.
(3) We further provide a query compiler from SPARQL into Spark SQL based on ExtVP that uses table statistics to select those tables with the highest selectivity for a given query. Our prototype called S2RDF is also available for download\footnote{\url{http://dbis.informatik.uni-freiburg.de/S2RDF}}.
(4) Finally, we present a comprehensive evaluation using the recent \textit{WatDiv} SPARQL diversity test suite~\cite{watdiv_2014} where we compare S2RDF with other state of the art SPARQL query engines for Hadoop and demonstrate its superior performance on very diverse query workloads.

\paragraph{Paper Structure}
In Sec.~\ref{sec:Foundations} we give a short introduction to RDF, SPARQL and Spark.
Sec.~\ref{sec:RelatedWork} summarizes related work on centralized and distributed RDF querying.
In Sec.~\ref{sec:RDFMapping} we provide an overview of the most common existing relational mappings for RDF before introducing our novel ExtVP schema based on semi-join reductions in Sec.~\ref{sec:ExtVP}.
Sec.~\ref{sec:Querying} describes the processing of SPARQL queries in S2RDF based on ExtVP and Sec.~\ref{sec:Evaluation} presents a comprehensive evaluation in comparison to state of the art approaches for distributed and centralized RDF querying using the WatDiv SPARQL diversity test suite.
Finally, we summarize the paper in Sec.~\ref{sec:Conclusion} and give an outlook on future work.

%% file: sections/foundations.tex
\subsection{RDF \& SPARQL}
\label{subsec:RDF_SPARQL}

\begin{figure}[htb]
	\centering
	\includegraphics[width=0.9\columnwidth]{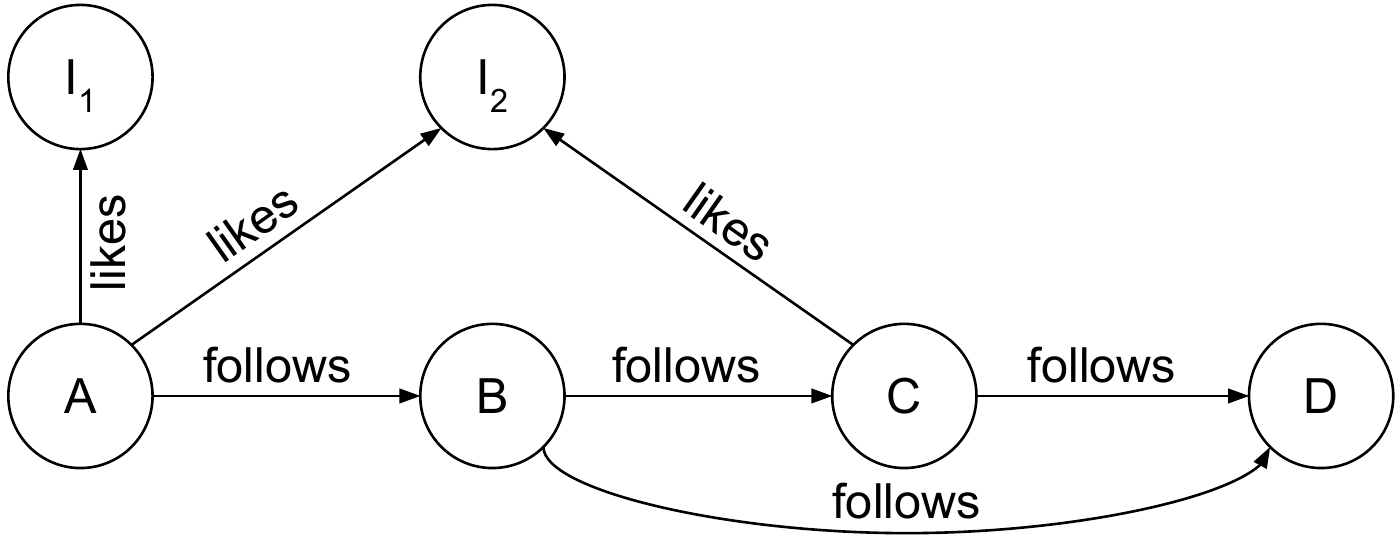}
	\caption{Visualization of RDF graph $G_1$}
	\label{fig:rdfgraph}
\end{figure}

RDF~\cite{rdfprimer} is the W3C recommended standard model for representing information about arbitrary resources. Global identifiers (\textit{IRIs}) are used to identify a resource, e.g.~the IRI for Leonardo da Vinci in DBpedia is \url{http://dbpedia.org/resource/Leonardo_da_Vinci}. For the sake of brevity, we use a simplified notation of RDF without IRIs in the following.
The basic notion of data modeling in RDF is a so-called \textit{triple} $t = (s,p,o)$ where $s$ is called \textit{subject}, $p$ \textit{predicate} and $o$ \textit{object}, respectively. It models the statement ``$s$ has property $p$ with value $o$" and can be interpreted as an edge from $s$ to $o$ labeled with $p$, $s \xrightarrow{p} o$. Hence, a set of triples forms a directed labeled (not necessarily connected) graph $G = \{ t_1, \ldots ,t_n \}$. For example, Fig.~\ref{fig:rdfgraph} visualizes the RDF graph
$G_1 = \{ (A, follows, B), (B, follows, C), (B, follows, D),$\\
$(C, follows, D), (A, likes, I_1), (A, likes, I_2), (C, likes, I_2) \}$.

\begin{figure}[htb]
	\centering
	\includegraphics[width=0.65\columnwidth]{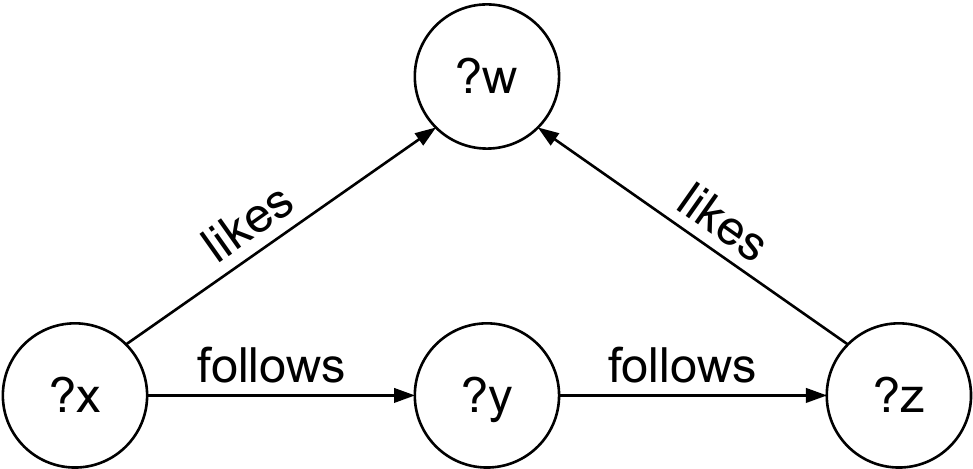}
	\caption{Query graph of $Q_1$}
	\label{fig:querygraph}
\end{figure}

SPARQL~\cite{sparql_10} is the W3C recommended query language for RDF. A SPARQL query $Q$ defines a graph pattern $P$ that is matched against an RDF graph $G$. This is done by replacing the variables in $P$ with elements of $G$ such that the resulting graph is contained in $G$ (pattern matching).
The basic notion in SPARQL is a so-called \textit{triple pattern} $tp = (s', p', o')$ with $s' \in \{ s,?s \}$, $p' \in \{ p,?p \}$ and $o' \in \{ o,?o \}$, i.e.~a triple where every part is either an RDF term (called \textit{bound}) or a variable (indicated by $?$ and called \textit{unbound}).
A set of triple patterns forms a \textit{basic graph pattern} (BGP).
For example, the following query $Q_1$ (in SPARQL syntax) contains a single BGP:

\begin{verbatim}
SELECT * WHERE {
  ?x likes ?w . ?x follows ?y .
  ?y follows ?z . ?z likes ?w
}
\end{verbatim}

\noindent
It can be interpreted as \textit{``For all users, determine the friends of their friends who like the same things"}. The query graph of $Q_1$ is illustrated in Fig.~\ref{fig:querygraph}.
Matched against RDF graph $G_1$ it gives a single result $(?x \rightarrow A, ?y \rightarrow B, ?z \rightarrow C, ?w \rightarrow I_2)$.
We use RDF graph $G_1$ and SPARQL query $Q_1$ as a running example throughout this paper.

The result of a BGP is a bag of so-called \textit{solution mappings} similar to relational tuples, i.e.~it does not produce triples and hence is not closed.
More formally, we can define the result of a BGP analogous to~\cite{perez_semantics_2009}:

Let $V$ be the infinite set of query variables and $T$ be the set of valid RDF terms.
A \textit{(solution) mapping} $\mu$ is a partial function $\mu:V \rightarrow T$. We call $\mu(?v)$ the variable binding of $\mu$ for $?v$ and $vars(tp)$ the set of variables contained in triple pattern $tp$. Abusing notation, for a triple pattern $tp$ we call $\mu(tp)$ the triple that is obtained by substituting the variables in $tp$ according to $\mu$. The \textit{domain} of $\mu$, $dom(\mu)$, is the subset of $V$ where $\mu$ is defined.

Two mappings $\mu_1, \mu_2$ are called \textit{compatible}, $\mu_1 \sim \mu_2$, iff for every variable $?v \in dom(\mu_1) \cap dom(\mu_2)$ it holds that $\mu_1(?v) = \mu_2(?v)$. It follows that mappings with disjoint domains are always compatible and the set-union (merge) of two compatible mappings, $\mu_1 \cup \mu_2$, is also a mapping.
The answer to a triple pattern $tp$ for an RDF graph $G$ is a bag of mappings $\Omega_{tp} = \lbrace \mu \mid dom(\mu) = vars(tp), \mu(tp) \in G \rbrace$.

The merge of two bags of mappings, $\Omega_1 \Join \Omega_2$, is defined as the merge of all compatible mappings in $\Omega_1$ and $\Omega_2$, i.e. $\Omega_1 \Join \Omega_2 = \lbrace (\mu_1 \cup \mu_2) \mid \mu_1 \in \Omega_1, \mu_2 \in \Omega_2, \mu_1 \sim \mu_2 \rbrace$. It can also be interpreted as a join on the variables that occur in both mappings.
Finally, the result to a basic graph pattern $bgp = \{ tp_1, \ldots, tp_m \}$ is defined as the merge of all mappings from all triple patterns, $\Omega_{bgp} = \Omega_{tp_1} \Join \ldots \Join \Omega_{tp_m}$.

On top of these basic patterns, SPARQL also provides more relational-style operators like \textsc{Optional} and \textsc{Filter} to further process and combine the resulting mappings.
This means that all other SPARQL operators use a bag of mappings as input, i.e.~they are not evaluated on the underlying RDF graph directly but on the result of one or more BGPs or other operators. Consequently, the most important aspect to query RDF data efficiently, is an efficient evaluation of BGPs.
For a detailed definition of the SPARQL syntax, we refer the interested reader to the official W3C Recommendation~\cite{sparql_10}. A formal definition of the SPARQL semantics can also be found in~\cite{perez_semantics_2009}.

\begin{figure}[htb]
	\centering
	\includegraphics[width=1.0\columnwidth]{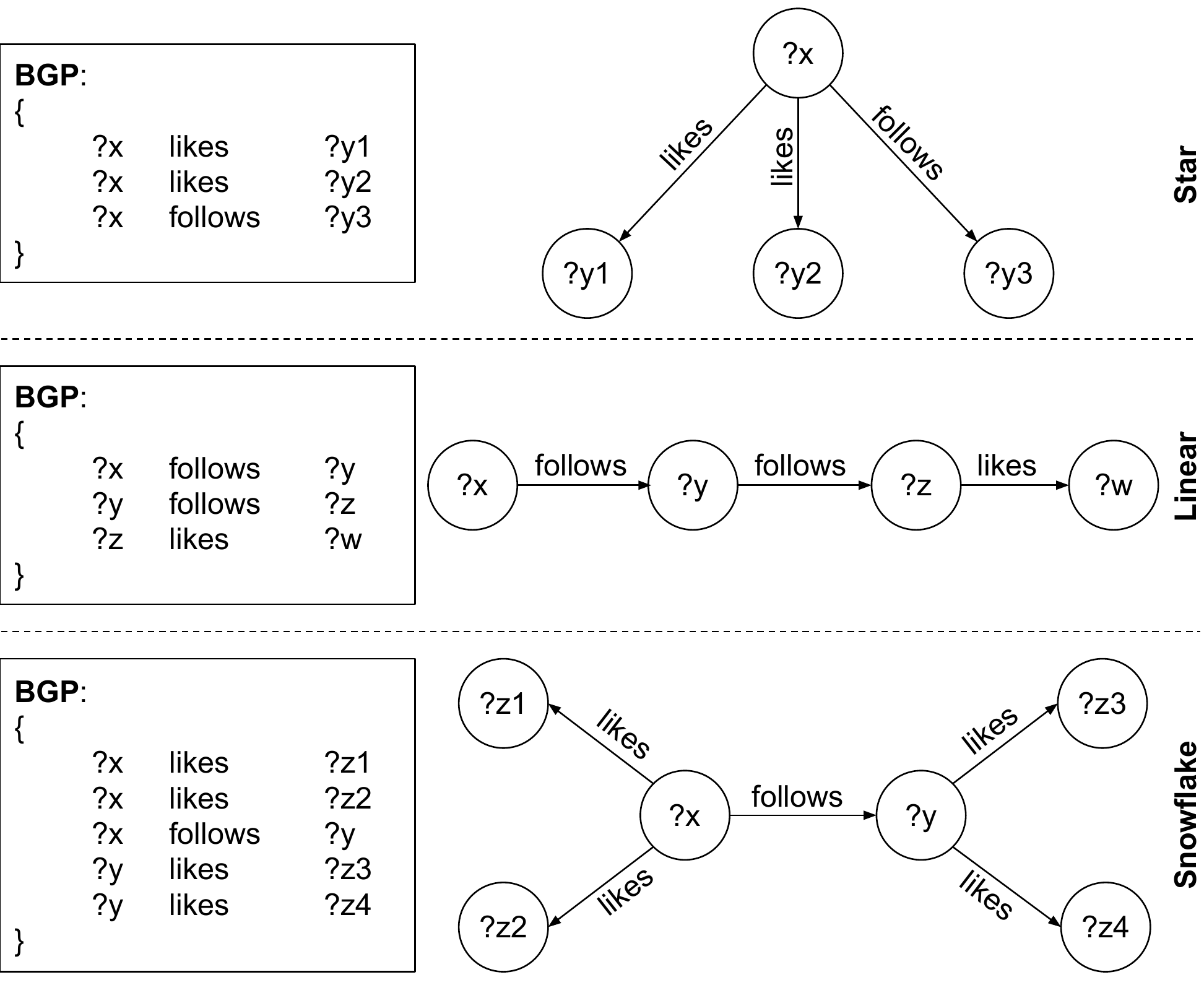}
	\caption{SPARQL BGP query shapes}
	\label{fig:queryshapes}
\end{figure} 

\paragraph{Query Shapes}
SPARQL BGPs can have different shapes depending on the position of variables in the triple patterns which can have severe impacts on query performance~\cite{watdiv_2014}. Fig.~\ref{fig:queryshapes} illustrates the most common BGP patterns.
The \textit{diameter} of a SPARQL BGP is defined as the longest path, i.e.~longest connected sequence of triple patterns, ignoring edge direction.
\textit{Star-shaped} patterns have a diameter of one and occur frequently in SPARQL queries, thus many query processors are optimized for this kind of workload. They are characterized by \textit{subject-subject} joins between triple patterns as the join variable is on subject position.
\textit{Linear-} or \textit{path-shaped} patterns are also very common in graph querying, e.g.~famous friend-of-a-friend queries. Linear patterns are made of \textit{object-subject} (or \textit{subject-object}) joins, i.e.~the join variable is on subject position in one triple pattern and on object position in the other. Thus, the diameter corresponds to the number of triple patterns. The performance of such workloads is often worse compared to star-shaped patterns in many RDF triplestores as the selectivity is typically lower and result sets can become large for highly connected graphs.
\textit{Snowflake-shaped} patterns are combinations of several star shapes connected by typically short paths. More complex query structures are essentially compositions of these fundamental patterns.

\subsection{Spark}
\label{subsec:Spark}

\textit{Spark}~\cite{Spark_2012} is a general-purpose in-memory cluster computing system that can run on Hadoop and process data from any Hadoop data source. The central data structure is a so-called \textit{\underline{R}esilient \underline{D}istributed \underline{D}ataset} (RDD)~\cite{RDD_2012} which is a fault-tolerant collection of elements that can be operated on in parallel. Spark attempts to keep an RDD in memory and partitions it across all machines in the cluster.
Conceptually, Spark adopts a \textit{data-parallel} computation model that builds upon a record-centric view of data, similar to \textit{MapReduce} and \textit{Apache Tez}. A job is modeled as a directed acyclic graph (DAG) of tasks where each task runs on a horizontal partition of the data.

Besides performance, the power of Spark comes from its flexibility that assembles previously disparate functionalities in a single unified data processing framework. It comes with a rich stack of high-level tools for batch processing, complex analytics, interactive exploration, graph and real-time stream processing as well as a relational interface called Spark SQL~\cite{Spark_SQL_2015}.
The cornerstone of Spark SQL is a so-called \textit{DataFrame} which is essentially a distributed collection of rows with the same schema and in that sense a special kind of RDD. They are equivalent to a table in a relational database. The advantage of DataFrames compared to classic RDDs is that the built-in optimizer of Spark SQL, \textit{Catalyst}, can apply more tailored optimizations since it knows the data schema and query semantics. When cached in memory, a DataFrame is represented using a columnar storage format that allows to access only required columns and reduces memory footprint by means of compression schemes such as dictionary and run-length encoding.

Spark SQL supports operating on a variety of Hadoop data sources through the DataFrame interface.
We use the general-purpose \textit{Parquet}\footnote{\url{https://parquet.apache.org/}} columnar storage format to persist the data store of S2RDF in HDFS due to several benefits.
First, the schema of tables is preserved such that Spark SQL can automatically infer it on-the-fly when loaded again.
Second, Parquet uses \textit{snappy}\footnote{\url{http://google.github.io/snappy/}} compression in combination with dictionary and run-length encoding to reduce storage consumption and boost access performance.
Third, as Parquet is a general-purpose format and not exclusively tied to Spark, Parquet files created by Spark SQL can also be used by other Hadoop frameworks, e.g.~we can directly load and query the data with \textit{Impala}~\cite{Impala_2015} just as well without any need for data movement or preparation.

%% file: sections/related_work.tex
In the last decade, a large variety of RDF storage and querying systems have been developed. A complete listing is out of scope for this work, thus we refer the interested reader to a couple of recent surveys~\cite{faye_survey_2012, kaoudi_rdf_2014, nitta_survey_2014, sakr_relational_2010}.
RDF stores can be broadly categorized in \textit{centralized} and \textit{distributed} systems, running on single machine or on a computing cluster (mostly using shared nothing architecture), respectively.
Additionally, they can be distinguished by their storage subsystem, i.e.~if they use some kind of relational database to store RDF data (\textit{relational-backed}), non-relational back-ends like key-value stores (\textit{NoSQL-backed}) or deploy an own storage subsystem tailored to RDF (\textit{native}).

\subsection{Centralized Systems}

Most early adopters used a centralized relational back-end to materialize RDF data using a giant \textit{triples table} or several \textit{property tables} (cf.~Sec.~\ref{sec:RDFMapping}), e.g.~\textit{Sesame}~\cite{broekstra_sesame_2002}, \textit{Jena}~\cite{carroll_jena_2004} and \textit{3store}~\cite{harris_3store_2003}. But also more recent state of the art systems like Virtuoso~\cite{erling_virtuoso_2010} and \textit{DB2RDF} (RDF support in DB2)~\cite{bornea_DB2RDF_2013} use a DBMS back-end.
In~\cite{abadi_vp_2007} Abadi et al.~proposed a \textit{vertically partitioned} approach using a two-column table for every unique predicate in an RDF dataset and describe an implementation of it in \cite{abadi_sw-store_2009} called \textit{SW-Store}. This approach also builds the basis of our ExtVP data layout in S2RDF that is explained in Sec.~\ref{sec:ExtVP}.

There also exists a bunch of centralized RDF systems that do not use a DBMS back-end but deploy their own RDF tailored storage solutions, e.g.~\cite{atre_BitMat_2010, harth_YARS_2005, rdf3x_2010, weiss_hexastore_2008, yuan_triplebit_2013, zou_gstore_2014}.
Most notably, \textit{RDF-3X}~\cite{rdf3x_2010} creates an exhaustive set of indexes for all RDF triple permutations and aggregated indexes for subsets, resulting in a total of 15 indexes stored in compressed clustered B+ trees.
\textit{Hexastore}~\cite{weiss_hexastore_2008} maintains six indexes for all triple permutations and can be seen as a combination of vertical partitioning \cite{abadi_vp_2007} and multiple indexing similar to~\cite{harth_YARS_2005}.
\textit{BitMat}~\cite{atre_BitMat_2010} uses a main-memory based bit-matrix structure to represent an RDF dataset. Each matrix corresponds to a slice along one dimension of a 3D bit-cube in which each cell is a bit representing the presence or absence of a unique RDF triple in the dataset.
\textit{TripleBit}~\cite{yuan_triplebit_2013} tries to address the scalability issues of other systems that rely on exhaustive indexes by designing a compact Triple Matrix that minimizes the number of indexes used in query evaluation and also reduces the size of intermediate results generated during query processing. Thus, it achieves comparable or better performance than other state of the art systems while using less storage and main memory.

\subsection{Distributed Systems}

In recent years, the ever-increasing size of RDF datasets has raised the need for distributed RDF storage and processing to overcome the inherent scalability limitations of centralized systems.
From a very general perspective, we can classify existing approaches to address this need in three broad categories: 

\paragraph{(1) Standalone Approaches}
The first type of systems are standalone distributed RDF stores, i.e.~they are self-contained and dedicated for distributed RDF processing, e.g.~\cite{boncz_VirtuosoCluster_2014, gurajada_TriAD_2014, harris_4store_2009, harth_yars2_2007, Clustered_TDB_2009, wylot_DiploCloud_2015}.

\textit{Virtuoso Cluster}~\cite{boncz_VirtuosoCluster_2014}, \textit{4store}~\cite{harris_4store_2009}, \textit{YARS2}~\cite{harth_yars2_2007} and \textit{Clustered TDB}~\cite{Clustered_TDB_2009} are extensions of the centralized systems Virtuoso~\cite{erling_virtuoso_2010}, 3store~\cite{harris_3store_2003}, YARS~\cite{harth_YARS_2005} and Jena~\cite{carroll_jena_2004} for distributed RDF processing, respectively.
\textit{TriAD}~\cite{gurajada_TriAD_2014} uses an asynchronous Message Passing protocol for distributed join execution in combination with join-ahead pruning via RDF graph summarization. RDF triples are distributed across a cluster of nodes using a locality-based, horizontal partitioning and stored in six in-memory triple vectors, each corresponding to one permutation of triple elements similar to Hexastore~\cite{weiss_hexastore_2008}. The METIS graph partitioner is used to partition the input RDF data graph and construct a summary graph. However, graph partitioning is an expensive task and centralized partitioner such as METIS are known to be limited in scalability \cite{lee_SHAPE_2013}. Thus, the initial centralized RDF partitioning can become a bottleneck with increasing data size.

\paragraph{(2) Federation Approaches}
The second type of systems use a federation of classical centralized RDF stores deployed on all cluster nodes and build a communication and coordination layer on top that distributes the data and (sub)queries, e.g.~\cite{galarraga_Partout_2014, hammoud_DREAM_2015, hose_WARP_2013, huang_HadoopRDF3X_2011, lee_SHAPE_2013, wu_Semstore_2014}. They mainly differ in the partitioning strategy used to spread the data across cluster nodes which also impacts the way how queries are split and executed. The general idea is that as much processing as possible is done locally at every node and a global aggregation mechanism merges the partial results, if necessary.

One of the first approaches in this direction that inspired most of the following work was introduced by Huang et al.~\cite{huang_HadoopRDF3X_2011}. It leverages the fact that RDF uses a graph data model by applying a graph partitioning algorithm such that triples which are nearby in the graph are stored on the same machine. They use METIS to do the partitioning and an instance of RDF-3X is used on all cluster nodes to store the allocated partitions and execute (sub)queries. By allowing partition borders to overlap, an $n$-hop guarantee assures that for a given vertex all vertices within a distance of at most $n$ are stored in the same partition. Thus, query patterns with a diameter of at most $n$ can be answered locally. If not, queries are decomposed and MapReduce is used to aggregate partial results from all cluster nodes. However, an $n$-hop guarantee imposes a large duplication of data in a densely connected RDF graph that growth exponentially with $n$ and hence only small $n$ are feasible (best results reported for $n = 2$). Similar to \cite{gurajada_TriAD_2014}, the initial partitioning suffers from the same scalability issues due to the use of a centralized graph partitioner and also performance degrades significantly when queries exceed the $n$-hop guarantee and hence MapReduce must be used for result aggregation.

\textit{Partout}~\cite{galarraga_Partout_2014} also uses RDF-3X on every cluster node but partitions RDF data with respect to a query workload such that queries can be processed by a minimum number of nodes in the cluster. First, data is fragmented based on an horizontal partitioning of triples defined by constant values in the triple patterns of the query workload. Second, derived fragments are allocated to nodes in the cluster such that as much queries as possible can be executed locally while maintaining a load balance. However, the typical query workload has to be known in advance and can also lead to suboptimal partitions if the workload changes over time.

\textit{SemStore}~\cite{wu_Semstore_2014} uses so-called Rooted Sub-Graphs (RSG) to partition RDF data where each RSG is placed on a single node in the cluster. In contrast to other partitioning strategies, RSGs are more coarse-grained and enable the system to localize more complex queries. A k-means partitioning algorithm implemented in MapReduce is used to place RSGs across the cluster such that data redundancy is reduced and possibility to execute queries locally is maximized. TripleBit is used on every cluster node to store and query data partitions allocated to that node.

The concept of \textit{DREAM}~\cite{hammoud_DREAM_2015} differs from the other systems in the sense that queries are partitioned instead of data. They argue that the choice of partitioning algorithm largely impacts the volume of intermediate results and is likely to produce suboptimal solutions for arbitrary query workloads. Hence, DREAM does not partition the data itself but all nodes in the cluster store a copy of the whole dataset which enables it to completely avoid intermediate data shuffling but only small auxiliary data has to be exchanged between cluster nodes. Of course, this implies a large duplication of data that increases with cluster size. SPARQL queries are partitioned and assigned to nodes based on a cost model and RDF-3X is used on every cluster node to store the entire dataset and execute the subparts of a query. The main drawback of DREAM is that the resources of a single node in the cluster can become a bottleneck as the whole dataset must be loaded into RDF-3X on every node. Thus, in terms of data scalability, DREAM provides no benefit compared to a centralized execution.

\paragraph{(3) Approaches based on Cloud Infrastructures}
The third type of systems are built on top of existing distributed platforms for Big Data processing, e.g.~\cite{husain_HadoopRDF_2011, papailiou_H2RDF+_2013, punnoose_Rya_2012, rohloff_SHARD_2011, pigsparql_2013, sempala_2014, zeng_Trinity.RDF_2013, zhang_EAGRE_2013}. 
Such platforms are also offered on a rental basis by leading Cloud providers, e.g.~\textit{Amazon Elastic Compute Cloud (EC2)}.
\textit{Hadoop} with its large surrounding ecosystem has become the most prominent and de facto industry standard for Big Data processing. The key concept is to have a common data pool (\textit{HDFS}) that can be accessed by various applications without the need for data duplication or movement.

\textit{SHARD}~\cite{rohloff_SHARD_2011} stores RDF data in HDFS grouped by subject and MapReduce is used for query processing using a so-called Clause-Iteration approach. A MapReduce job is created for every triple pattern of a query (called clause) which conceptually leads to a left-deep query plan.

\textit{PigSPARQL}~\cite{pigsparql_2013} also stores RDF directly in HDFS using a vertical partitioning schema. Instead of compiling SPARQL queries directly into MapReduce jobs, as done by SHARD, it uses \textit{Pig} as an intermediate layer. A query gets translated into Pig Latin, the programming language of Pig, which is then mapped to a corresponding sequence of MapReduce jobs. By using this two-level abstraction, PigSPARQL profits from sophisticated optimizations of Pig and runs on all platforms supported by Pig, including all versions of Hadoop.
Yet, as PigSPARQL and SHARD are both based on MapReduce for query execution, they suffer from relatively high query latencies. 

\textit{Rya}~\cite{punnoose_Rya_2012} stores its data in \textit{Accumulo}, a variant of Google's BigTable which is essentially a sorted and column-oriented NoSQL key-value store on top of HDFS. They use three tables SPO, POS and OSP (names correspond to order of triple elements) where triples are completely stored in the Row ID part of the key which effectively corresponds to three clustered indexes. As these tables are sorted by Row ID, each triple pattern can be answered by a range scan over one of these tables. Queries are executed using an index nested loop join locally at the server. This can be a potential limiting factor for scalability as join processing itself is not distributed, especially for non-selective queries with large intermediate result sets.

\textit{H2RDF+}~\cite{papailiou_H2RDF+_2013} is conceptually related to Rya as it stores RDF data in \textit{HBase} which is also a variant of Google's BigTable on top of HDFS. It also stores triples completely in the row key but uses six tables for all possible triple permutations thus creates six clustered indexes. Additionally, it also maintains aggregated index statistics to estimate triple pattern selectivity as well as join output size and cost. Based on these estimations, the system adaptively decides whether queries are executed centralized over a single cluster node or distributed via MapReduce. H2RDF+ comes with implementations of merge and sort-merge joins both for MapReduce and local execution. By combining both approaches, H2RDF+ can answer selective queries very efficiently while being able to scale for non-selective queries that cannot be answered efficiently with centralized execution. However, distributed query execution can be orders of magnitude slower than centralized.

\textit{Sempala}~\cite{sempala_2014} is conceptually related to S2RDF as it is also a SPARQL-over-SQL approach based on Hadoop. It is built on top of \textit{Impala}, a massive parallel processing (MPP) SQL query engine, while data is stored in HDFS using the \textit{Parquet} columnar storage format. The data layout of Sempala consists of a single \textit{unified property table} (cf.~Sec.~\ref{subsec:PropertyTables}) such that star-shaped queries can be answered without the need for a join. Hence, in contrast to ExtVP, this layout is targeted on a specific query shape. Complex queries are decomposed into disjoint triple groups (star-shaped subpatterns) that are joined to build the final result.

%% file: sections/rdf_mapping.tex
\subsection{Triples Table}
\label{subsec:TriplesTable}

It is a common approach by many RDF triplestores to manage RDF data in a relational DBMS back-end, e.g.~\cite{jena_2003}. These solutions typically use a giant so-called \textit{triples table} with three columns, containing one row for each RDF statement, i.e.~$TT(s, p, o)$.
Formally, a triples table over an RDF graph $G$ can be defined as:
\begin{tabbing}
$TT[G] = \{ (s,p,o) \in G \}$
\end{tabbing}

\begin{figure}[htb]
	\centering
	\includegraphics[width=1.0\columnwidth]{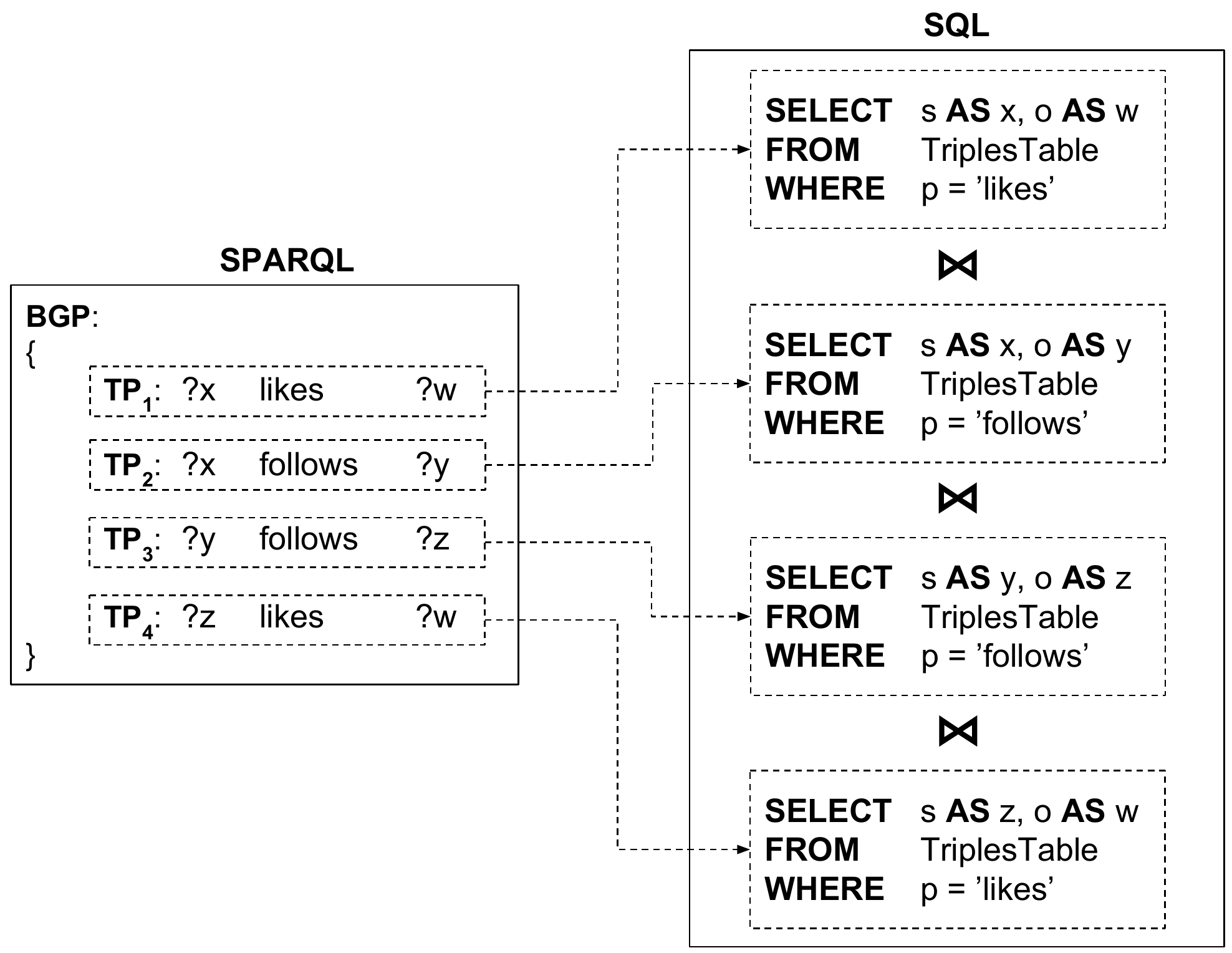}
	\caption{SPARQL to SQL for $Q_1$ based on TT}
	\label{fig:tripletable_sql}
\end{figure}

While being flexible and simple in its representation, it is not an efficient approach for large-scale datasets as query evaluation essentially boils down to a series of self-joins on this table. Fig.~\ref{fig:tripletable_sql} illustrates the mapping from SPARQL query $Q_1$ to SQL based on a triples table representation of RDF. As we can see, the table is referenced four times and different selections are used to retrieve the triples matching the corresponding triple pattern. A naive evaluation would result in four scans over the whole triples table. Of course, an optimizer can reduce it to a single scan but still this is far away from optimal as the whole dataset has to be touched at least once, even if the query only selects a very small subset.
Therefore, it is often accompanied by several indexes over some or all (six) triple permutations~\cite{rdf3x_2010, weiss_hexastore_2008}, e.g.~based on $B^+$-trees, for query speedup.

The triples table approach is conceptually the most straight forward representation of RDF in a relational model but it has several drawbacks in a distributed environment. Most notably, rich indexes are hardly supported by most frameworks on Hadoop due to their complexity in a distributed storage and processing environment which makes them hard to maintain. As a consequence, using a triples table typically requires a query processor based on Hadoop to scan the whole dataset and causes a lot of (unnecessary) network I/O to exchange data between cluster nodes. In our practical experience otherwise, we have observed that query optimization in an Hadoop setup basically always means to minimize I/O, especially from a network perspective. Hence, a triples table is suited for one-time processing of RDF data (e.g.~ETL) but not efficient for repeated querying.

\subsection{Vertical Partitioning}
\label{subsec:VerticalPartitioning}

An often used optimization is a \textit{vertical partitioned} (VP) schema, introduced by Abadi et al.~in~\cite{abadi_vp_2007}. Instead of a single three-column table, it uses a two-column table for every RDF predicate, e.g.~$follows(s, o)$. It evolved from the observation that a triple pattern in a SPARQL query is typically predicate-bound and hence all triple candidates matching the pattern can be found in the corresponding VP table, leading to a significant reduction of query input size.
Let $\mathcal{P}$ be the set of all predicates in an RDF graph $G$. Formally, a vertical partitioning over $G$ can be defined as:
\begin{tabbing}
$VP_p[G]$ \= $= \{ (s,o) \mid (s,p,o) \in G \}$ \\[1ex]
$VP[G]$ \> $= \{ VP_p[G] \mid p \in \mathcal{P} \}$
\end{tabbing}

\begin{figure}[htb]
	\centering
	\includegraphics[width=1.0\columnwidth]{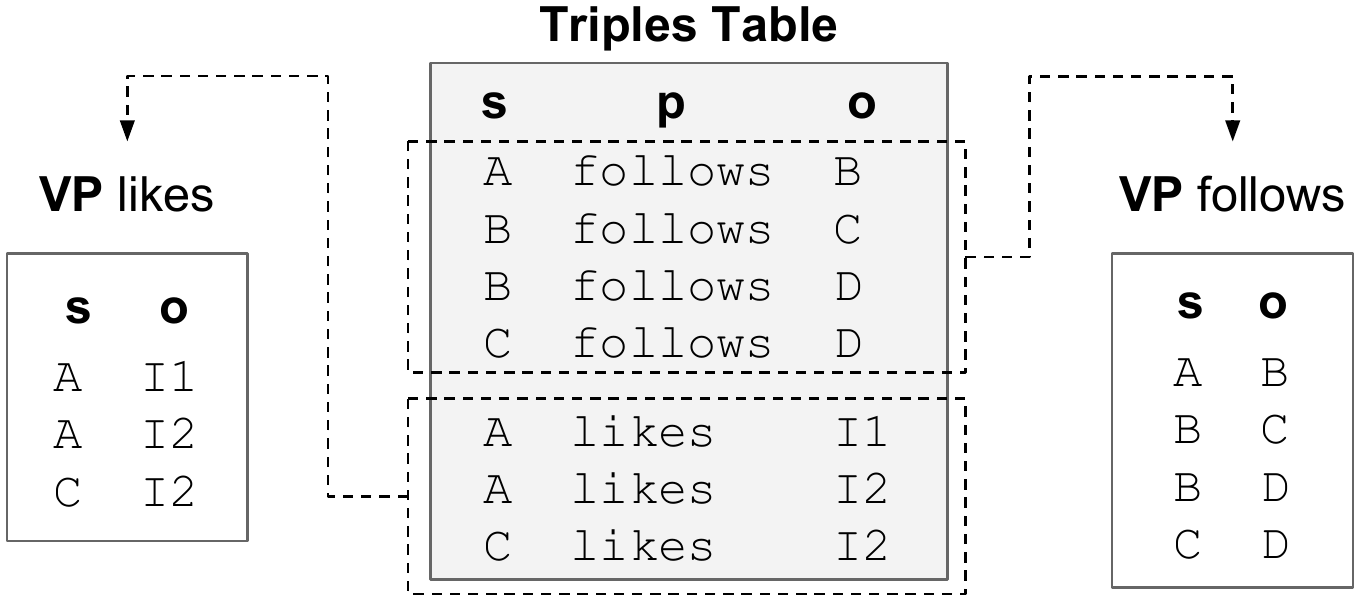}
	\caption{Vertical Partitioning of $G_1$}
	\label{fig:tripletable_vp}
\end{figure}

\begin{figure}[htb]
	\centering
	\includegraphics[width=1.0\columnwidth]{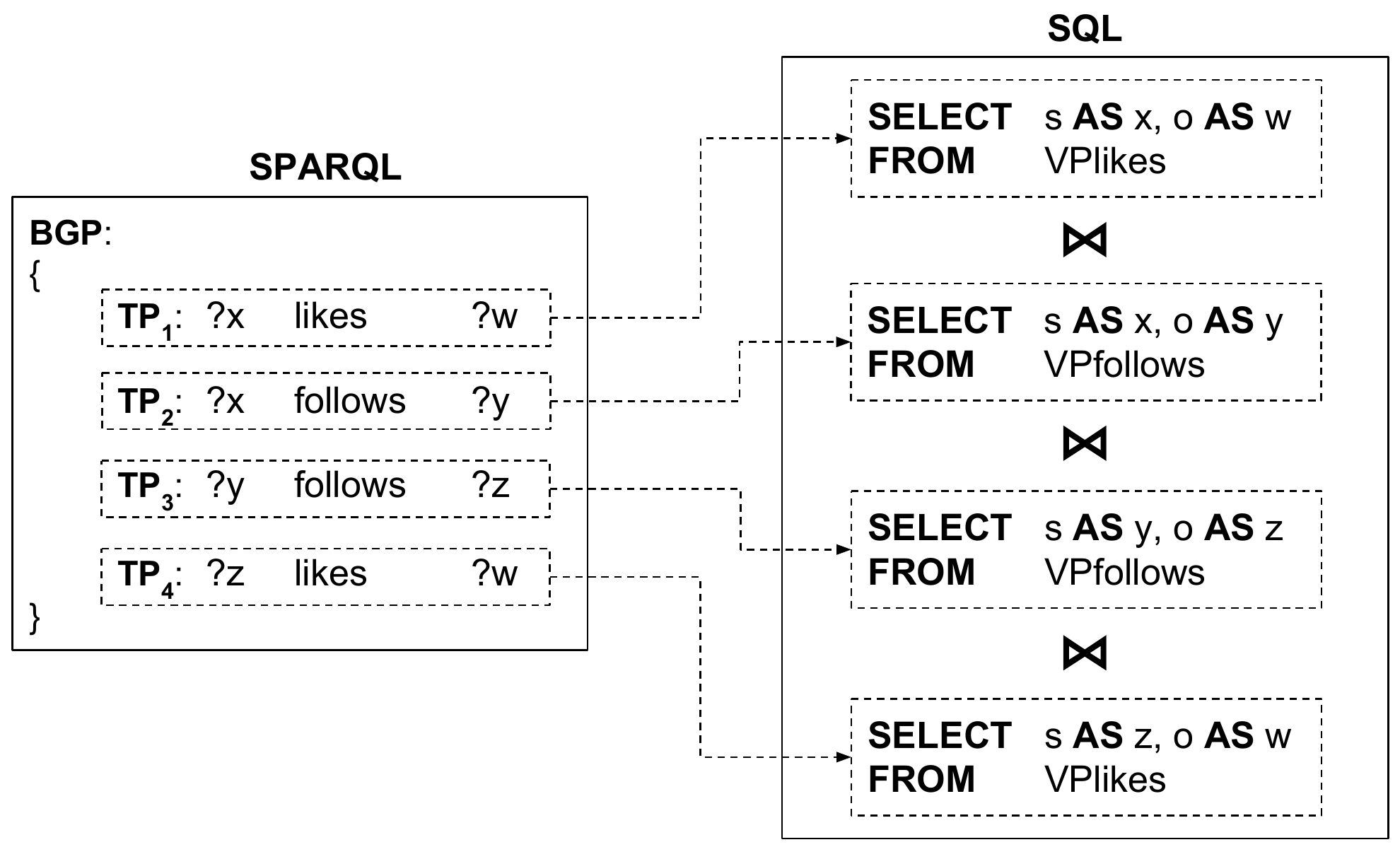}
	\caption{SPARQL to SQL for $Q_1$ based on VP}
	\label{fig:vp_sql}
\end{figure}

For RDF graph $G_1$, the vertical partitioning is composed of two tables for predicates \textit{follows} and \textit{likes} (see~Fig.~\ref{fig:tripletable_vp}). As a consequence, the SPARQL to SQL mapping can now select the corresponding VP tables for each triple pattern without the need for additional selections (cf.~Fig.~\ref{fig:vp_sql}).

For large RDF datasets with many predicates, this representation proves to be efficient in practice as it mimics the effect of an index on predicates and is also easy to manage in a distributed Hadoop environment. However, some partitions can account for a large portion of the entire graph and cause a lot of I/O. Especially for selective queries, there is still a lot of unnecessary shuffling of data involved that gets discarded afterwards.

\subsection{Property Tables}
\label{subsec:PropertyTables}

Other representations try to reduce the number of joins needed for BGP evaluation. A typical approach in this direction are so-called \textit{property tables}~\cite{jena_2006} where all predicates (or properties) that tend to be used in combination are stored in one table, e.g.~all predicates used to describe a person.
Formally, a property table over an RDF graph $G$ for predicates $p_1, \ldots, p_n$ can be defined as:
\begin{tabbing}
$PT_{p_1, \ldots, p_n}[G] = \{ (s,o_1,\ldots,o_n) \mid$ \= $(s,p_1,o_1) \in G \wedge \ldots \wedge$ \\[0.5ex]
\> $(s,p_n,o_n) \in G \}$
\end{tabbing}

The columns (predicates) grouped together in a property table are either determined by a clustering algorithm or by type definitions in the dataset itself. The weakness of this approach is the existence of multi-valued predicates in RDF, e.g.~a user typically follows more than one other user. Such predicates are either modeled using additional auxiliary tables or by entry duplication~\cite{sempala_2014}. A property table $PT_{follows,likes}$ for example graph $G_1$ using a row duplication strategy for multi-valued predicates is given in Table~\ref{tab:propertyTable}.

\begin{table}[htb]
\caption{Property table for RDF graph $G_1$}
\label{tab:propertyTable}
\scriptsize
\begin{tabularx}{\columnwidth}{XXX}
\toprule
\textbf{s} & \textbf{follows}:string & \textbf{likes}:string\\\midrule
A & B & I$_1$ \\
A & B & I$_2$ \\
B & C & \\
B & D & \\
C & D & I$_2$ \\\bottomrule
\end{tabularx}
\end{table}

\begin{figure}[htb]
	\centering
	\includegraphics[width=1.0\columnwidth]{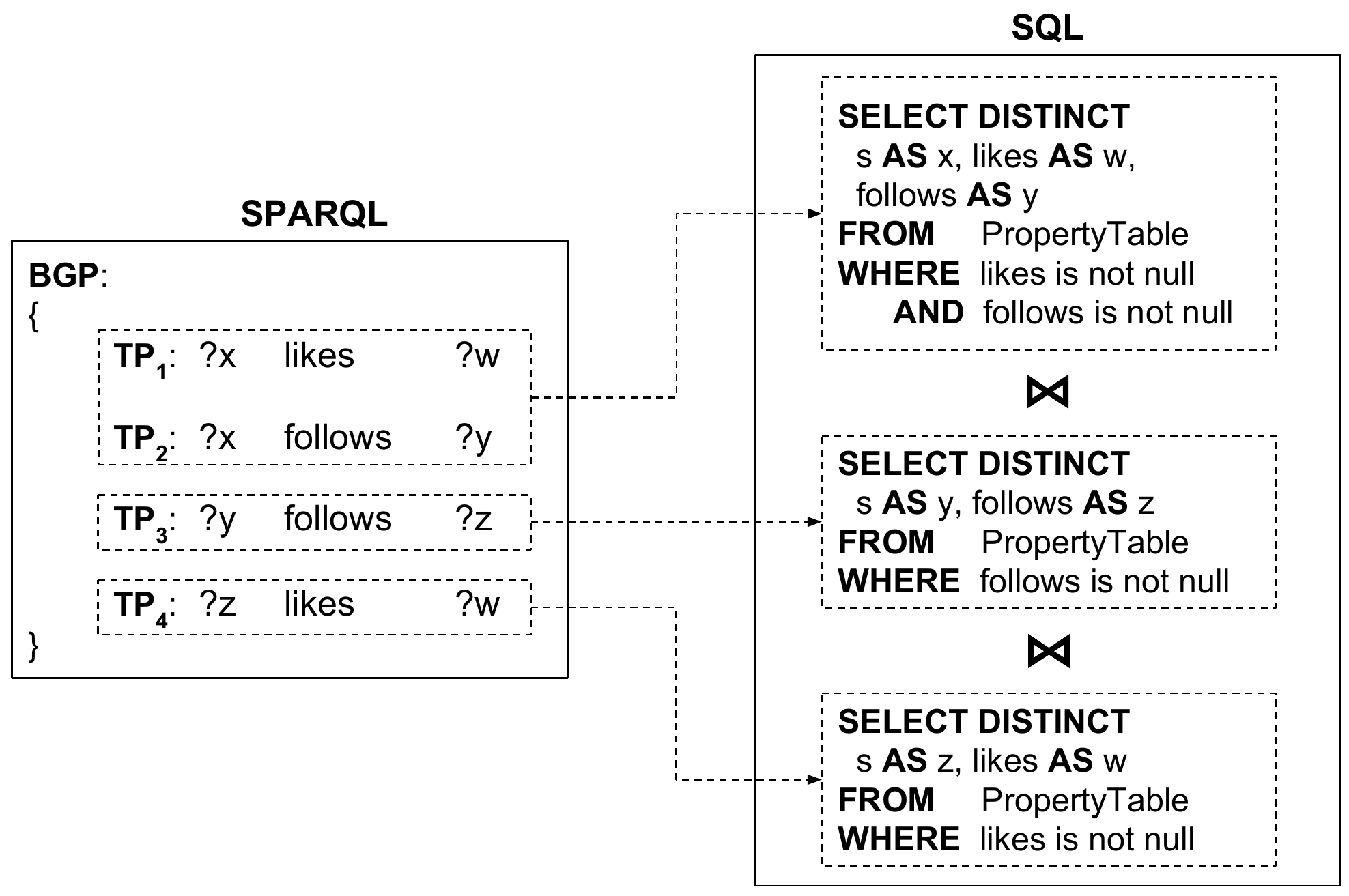}
	\caption{SPARQL to SQL for $Q_1$ based on PT}
	\label{fig:propertytable_sql}
\end{figure}

The biggest advantage of property tables compared to a triples table or VP approach is the reduction of subject-subject self-joins for star-shaped query patterns. For example, consider the first two triple patterns of query $Q_1$. They share the same variable on subject position and hence form a star pattern (cf.~Fig.~\ref{fig:querygraph}) that can be answered without the need for a join using a property table. Consequently, the SPARQL to SQL mapping for query $Q_1$ based on a property table representation needs only two joins instead of three compared to triples table and VP (see Fig.~\ref{fig:propertytable_sql}).
While star-shaped queries are first-class citizens in a property table approach, it neglects other query patterns. Especially for linear-shaped patterns, a property table has no real benefit compared to a simple triples table or can be even worse in practice.

%% file: sections/extvp.tex
Besides general query optimization techniques, data layout plays an important role for efficient SPARQL query evaluation in a distributed environment. Due to the lack of rich indexes in contrast to traditional single machine databases, a carelessly chosen data model can cause severe performance issues for some query shapes. Regarding the efficient evaluation of SPARQL BGPs in an Hadoop setting, one can conceptually distinguish two design goals: (1) the reduction of input data size and thus I/O in general, and (2) the saving of join operations. Vertical partitioning (cf.~Sec.~\ref{subsec:VerticalPartitioning}) and property tables (cf.~Sec.~\ref{subsec:PropertyTables}) are prominent examples for the first or the latter design goal, respectively.

Furthermore, one has also to consider the specific properties of the underlying execution layer, in our case Spark. In an initial pre-evaluation, we therefore examined the main influence factors of query performance in Spark SQL. The most important finding was that data input size reduction is much more effective than the saving of join operations. We attribute this to the fact that Spark is an in-memory system optimized for pipelined execution with little to no setup overhead for individual operations. Not surprisingly, our workload is much more I/O than CPU bound and especially network I/O is the most limiting factor.

The figures in the following use our running example introduced in Sec.~\ref{subsec:RDF_SPARQL}.

\subsection{ExtVP in a Nutshell}
\label{subsec:ExtVPNutshell}

Many existing data layouts for RDF are tailored towards a specific kind of query shape (cf.~Sec.~\ref{subsec:RDF_SPARQL}) and most often star-shaped queries are the primary focus as these shapes occur very often in typical SPARQL queries, especially in rather selective ones. Property tables (cf.~Sec.~\ref{subsec:PropertyTables}) are a prominent example for this. One of the design criteria for the data layout of S2RDF is that it should not favor any specific query shape and in particular also provide an improvement for linear (path) patterns as they are often neglected by existing approaches. Furthermore, it should also be applicable for queries with a large diameter as well as complex mixed shapes being a composition of star, linear and snowflake patterns.

Based on our pre-evaluation findings, we decided to use a vertical partitioned (VP) schema as the base data layout for RDF in S2RDF. Using such a schema, the results for a triple pattern with bound predicate can be retrieved by accessing the corresponding VP table only which leads to a large reduction of the input size, in general. Unfortunately, the size of these tables is highly skewed in a typical RDF dataset with some tables containing only a few entries while others comprise a large portion of the entire graph. If a query needs to access such a table, there is still a lot of data to be read (and potentially shuffled) that gets discarded in the following when joined with the results of other triple patterns. To improve this behavior, we developed an extension to the VP schema called \textit{\underline{Ext}ended \underline{V}ertical \underline{P}artitioning} (ExtVP).

\begin{figure}[htb]
	\centering
	\includegraphics[width=1.0\columnwidth]{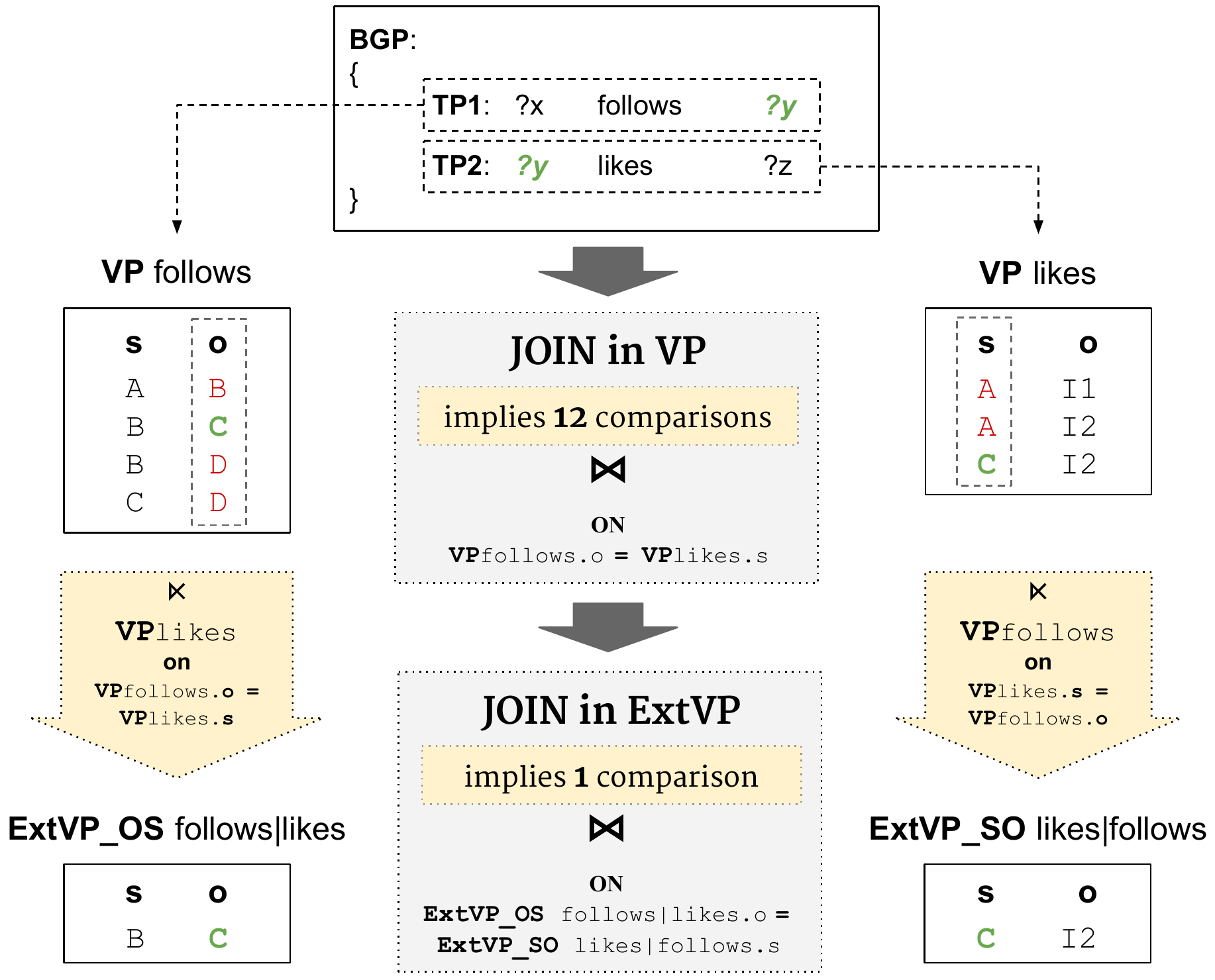}
	\caption{Join comparisons in VP and ExtVP}
	\label{fig:join_vp_extvp}
\end{figure}

The basic idea of ExtVP is to precompute the possible join relations between partitions (i.e.~tables) of VP. For example, consider the BGP in Fig.~\ref{fig:join_vp_extvp} consisting of two triple patterns and the corresponding VP tables for predicates \textit{follows} and \textit{likes} of RDF graph $G_1$. To compute the final result using VP, both tables are joined depending on the position of variable $?y$. This implies 12 comparisons in total but there is only one matching tuple in both tables. Furthermore, it requires both tables to be read and at least partially shuffled, although most tuples will not find a corresponding join partner (\textit{dangling} tuples).

In ExtVP, the query processor can directly access the subset of table \textit{follows} where the object also exist as a subject in at least one tuple in table \textit{likes} and join it with the equivalent subset of table \textit{likes}. This avoids dangling tuples to be used as input and thus also reduces I/O. In addition, it also reduces the number of comparisons in join computation which further speeds up the overall processing. For example, in Fig.~\ref{fig:join_vp_extvp} the number of comparisons is minimized to one when using ExtVP.
These extra join relations between VP tables can be precomputed and materialized by a series of semi-joins. In practice, the space requirement of ExtVP is reasonable and comparable to existing approaches, e.g.~\cite{papailiou_H2RDF+_2013, sempala_2014, weiss_hexastore_2008}. The size and overhead of ExtVP compared to VP is discussed in more detail in Sec.~\ref{subsec:ExtVPSize}.

\subsection{ExtVP Definition}
\label{subsec:ExtVPDefinition}

\begin{figure}[htb]
	\centering
	\includegraphics[width=1.0\columnwidth]{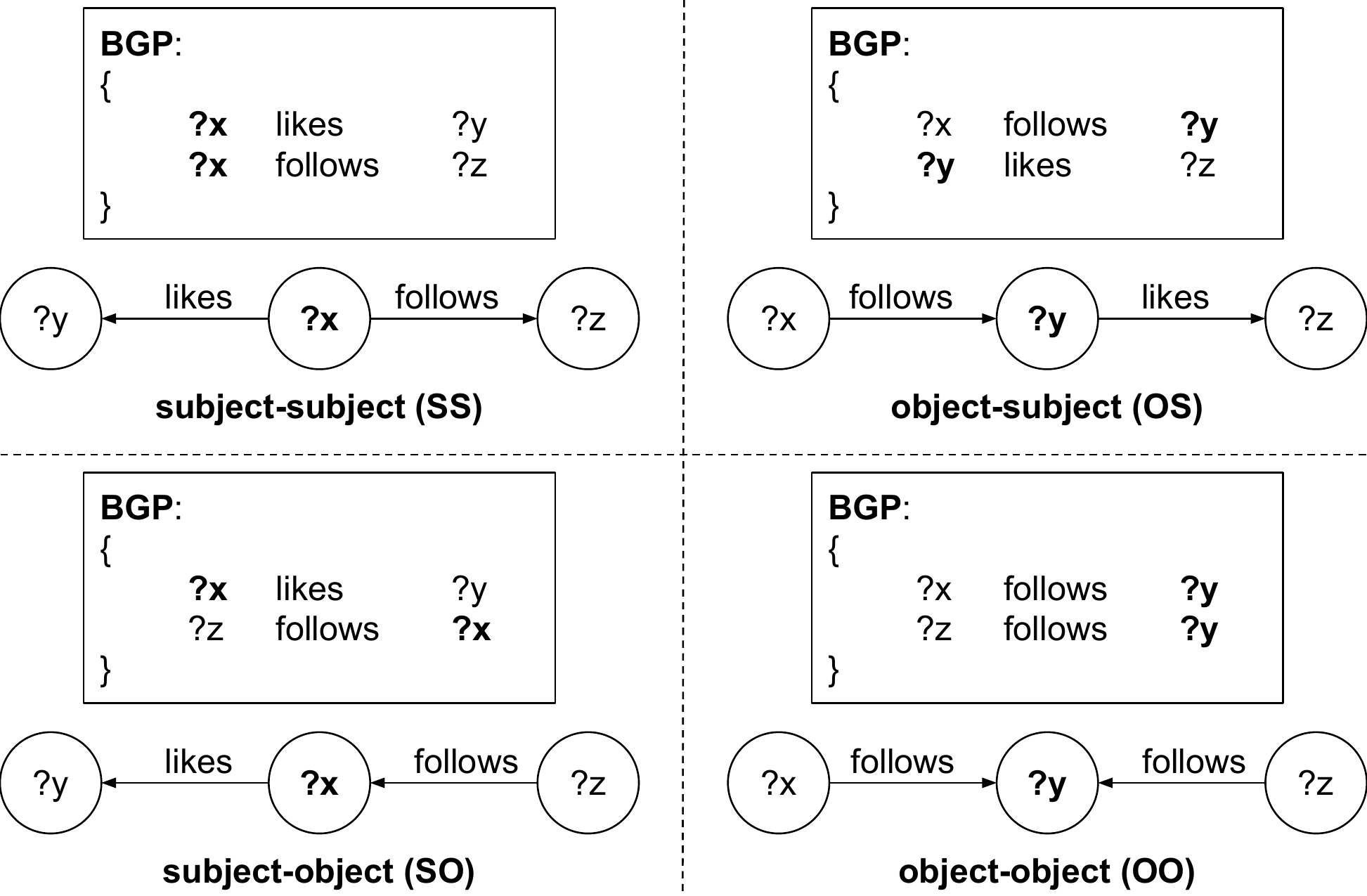}
	\caption{Correlations between triple patterns}
	\label{fig:join_positions}
\end{figure}

The relevant semi-joins between tables in VP are determined by the possible joins that can occur when combining the results of triple patterns during query execution.
The position of a variable that occurs in both triple patterns (called join variable) determines the columns on which the corresponding VP tables must be joined. We call the co-occurrence of a variable in two triple patterns a \textit{correlation}. Fig.~\ref{fig:join_positions} illustrates the possible correlations, e.g.~if the join variable is on subject position in both triple patterns we call this a \textit{subject-subject correlation} (SS) as both VP tables must be joined on their subjects. The other correlations are \textit{subject-object} (SO), \textit{object-subject} (OS) and \textit{object-object} (OO). We do not consider join variables on predicate position as such patterns are primarily used for inference or schema exploration but rarely used in a typical SPARQL query~\cite{abadi_vp_2007}. S2RDF can answer such queries by accessing the base triples table for triple patterns with unbound predicate but is not further optimized for it.

The goal of ExtVP is to avoid dangling tuples in join input tables, i.e.~tuples that do not find a join partner. They cause unnecessary I/O and comparisons during join execution as well as an increase in memory consumption. Since Spark is an in-memory system and memory is typically much more limited than HDFS disk space, saving this resource is important for scalability.
To this end, we determine the subsets of a VP table $VP_{p_1}$ that are guaranteed to find at least one match when joined with another VP table $VP_{p_2}$, $p_1, p_2 \in \mathcal{P}$. This can be done with a series of semi-joins.
We do this for SS, OS and SO correlations between all pairs of VP tables (i.e.~all pairs of predicates). We do not precompute OO correlations between VP tables as this would not help much in practice. Two triple patterns in a SPARQL query that are connected by an OO correlation often use the same predicates (cf.~OO in Fig.~\ref{fig:join_positions}) and thus result in a self-join of the corresponding VP table. In this case, a semi-join would reduce nothing but simply return the table itself. So we decided to not precompute OO correlations at all due to their relatively poor cost-benefit ratio. Indeed, it is only a design choice and we could precompute them just as well. The advantage of ExtVP is that none of these precomputations are mandatory. S2RDF makes use of it, if they exist, or uses the normal VP tables instead.
Furthermore, an optional selectivity threshold for ExtVP can be specified such that only those tables are materialized where reduction of the original VP tables is large enough. This reduces the size overhead of ExtVP and we discuss it in more detail in Sec.~\ref{subsec:ExtVPSize}.

For two VP tables $VP_{p_1}$ and $VP_{p_2}$ we compute the following semi-join reductions and materialize the results as separate tables in HDFS (if not empty and selectivity is within the threshold):
\begin{tabbing}
\textbf{SS:}~~~ \= $VP_{p_1} \ltimes_{s=s} VP_{p_2}$~,~ \= $VP_{p_2} \ltimes_{s=s} VP_{p_1}$\\
\textbf{OS:} \> $VP_{p_1} \ltimes_{o=s} VP_{p_2}$~,~ \> $VP_{p_2} \ltimes_{o=s} VP_{p_1}$\\
\textbf{SO:} \> $VP_{p_1} \ltimes_{s=o} VP_{p_2}$~,~ \> $VP_{p_2} \ltimes_{s=o} VP_{p_1}$
\end{tabbing}

Essentially, the idea of ExtVP comes from the fact that a join between two tables $T_1, T_2$ on attributes $A, B$ can be decomposed in the following way:\\
\centerline{$T_1 \Join_{A=B} T_2 = (T_1 \ltimes_{A=B} T_2) \Join_{A=B} (T_1 \rtimes_{A=B} T_2)$}\\
This is a common join optimization technique in distributed database systems to reduce communication costs~\cite{ozsu_principles_2011}. We do not precompute the actual join results itself as this usually would increase space consumption by an order of magnitude or more. Semi-join reductions on the other side are always guaranteed to be a subset of the corresponding base table.
Formally, an ExtVP schema over an RDF graph $G$ can be defined as:

\begin{tabbing}
$ExtVP_{p_1|p_2}^{SS}[G]$ \= $= \{ (s,o) \mid (s,o) \in VP_{p_1}[G] \wedge$\\
\> $\qquad \qquad \exists (s',o') \in VP_{p_2}[G]: s = s' \}$\\[0.5ex]
\> $\equiv VP_{p_1}[G] \ltimes_{s=s} VP_{p_2}[G]$\\[2ex]
$ExtVP^{SS}[G]$ \> $= \{ ExtVP_{p_1|p_2}^{SS}[G] \mid p_1, p_2 \in \mathcal{P} \wedge p_1 \neq p_2 \}$\\
\\[1.5ex]
$ExtVP_{p_1|p_2}^{OS}[G]$ \= $= \{ (s,o) \mid (s,o) \in VP_{p_1}[G] \wedge$\\
\> $\qquad \qquad \exists (s',o') \in VP_{p_2}[G]: o = s' \}$\\[0.5ex]
\> $\equiv VP_{p_1}[G] \ltimes_{o=s} VP_{p_2}[G]$\\[2ex]
$ExtVP^{OS}[G]$ \> $= \{ ExtVP_{p_1|p_2}^{OS}[G] \mid p_1, p_2 \in \mathcal{P} \}$\\
\\[1.5ex]
$ExtVP_{p_1|p_2}^{SO}[G]$ \= $= \{ (s,o) \mid (s,o) \in VP_{p_1}[G] \wedge$\\
\> $\qquad \qquad \exists (s',o') \in VP_{p_2}[G]: s = o' \}$\\[0.5ex]
\> $\equiv VP_{p_1}[G] \ltimes_{s=o} VP_{p_2}[G]$\\[2ex]
$ExtVP^{SO}[G]$ \> $= \{ ExtVP_{p_1|p_2}^{SO}[G] \mid p_1, p_2 \in \mathcal{P} \}$\\
\\[1.5ex]
$ExtVP[G] = \{ ExtVP^{SS}[G], ExtVP^{OS}[G], ExtVP^{SO}[G] \}$
\end{tabbing}

\begin{figure*}[htb]
	\centering
	\includegraphics[width=1.0\textwidth]{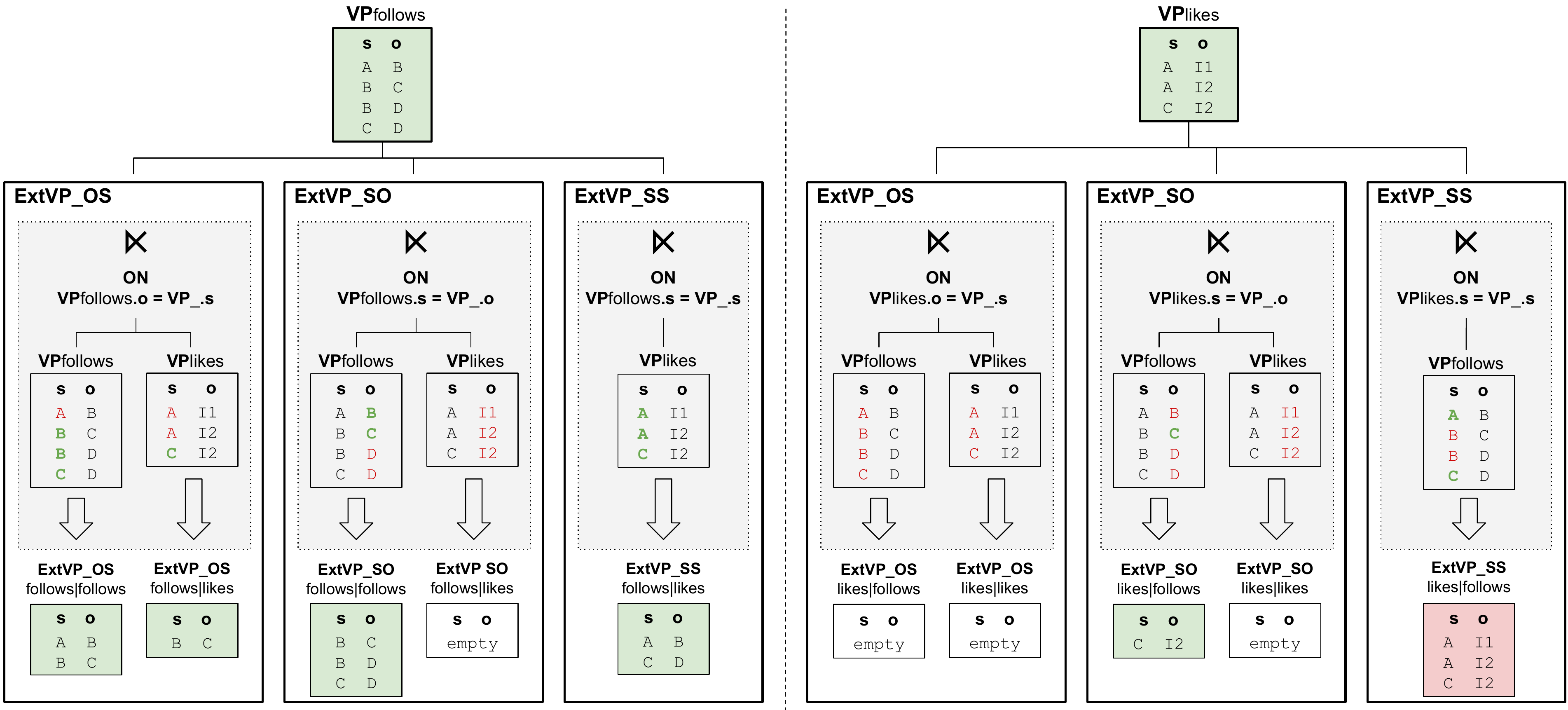}
	\caption{ExtVP data model for RDF graph $G_1$. Left side shows ExtVP tables derived from $VP_{follows}$, right side for $VP_{likes}$, resp. Tables marked in green are stored. Red tables are equal to VP and thus not stored.}
	\label{fig:extvp_all}
\end{figure*}

The benefit of ExtVP depends on the magnitude of reduction compared to the base VP tables.
Fig.~\ref{fig:extvp_all} illustrates the entire ExtVP schema for RDF graph $G_1$ consisting of two predicates. The green-marked tables are actually stored in HDFS and used for query processing. These are the VP tables itself and those correlation tables of ExtVP that give a reduction of the corresponding VP table. Empty tables and those that do not give a reduction (red-marked) are not stored. In a typical heterogeneous RDF dataset consisting of several classes, many ExtVP tables would be empty because their predicates cannot be combined (e.g.~users have different predicates than products). To avoid these unnecessary semi-join operations when constructing an ExtVP schema, we determine those predicates that have any correlation to one another. For example, the following query determines those predicates that have an OS correlation to predicate~$p_1$:
\\
\verb|SELECT DISTINCT TT.p FROM TriplesTable TT|\\
\verb|LEFT SEMI JOIN VPp1 ON TT.o = VPp1.s|

\paragraph{ExtVP and Classical Database Query Optimization}
The notion of semi-joins was originally proposed to reduce communication cost for join processing in early distributed databases~\cite{bernstein_semijoin_1981}. The idea is to apply semi-joins on the fly during query processing such that a relation is reduced before shipped to another site for join processing. The goal is to derive a semi-join program, i.e.~a sequence of semi-joins, such that no dangling tuples in any relation are shipped when processing a sequence of join operations. Such a program is called a \textit{full reducer} but its derivation is NP-hard, in general.

However, query processing in Spark works different in the sense that a relation is not stored on one site and shipped to another site but joins are executed in parallel on all cluster nodes on portions of the data, similar to an MPP (Massively Parallel Processing) database. This makes the application of semi-joins on the fly during query processing less effective.
Furthermore, VP tables are very narrow with only two columns which further limits the benefits of semi-joins during query processing. Our experiments have shown that on the fly application of semi-joins on VP tables during query processing gives no real performance benefit compared to standard join execution in Spark on average.
But as all VP tables have only two columns, we can afford to precompute semi-join reductions of these tables for all possible correlations (omitting OO correlations for aforementioned reasons). This way, we do not have to compute them on the fly but only once in advance.

Conceptually, the idea of ExtVP is related to the notion of \textit{Join Indices}~\cite{valduriez_JoinIndices_1987} in relational databases and \textit{Access Support Relations}~\cite{kemper_ASR_1992} in object-oriented databases. An $ExtVP_{p_1|p_2}$ table basically resembles a \textit{clustered join index} between VP tables $VP_{p_1}$ and $VP_{p_2}$ as we store the actual table payload in the index instead of unique surrogates. We can afford this because VP tables, in contrast to an arbitrary relational schema, have a fixed two-column layout.
Access support relations (ASR) have been introduced to facilitate path expression evaluation. In principle, $ExtVP^{OS}$ resembles a binary decomposition of all possible ASR (i.e.~paths) in an RDF graph following the edges in forward direction. In addition, $ExtVP^{SO}$ can be seen as a binary decomposition following the edges in backward direction.

\subsection{ExtVP Selectivity Threshold}
\label{subsec:ExtVPSize}

ExtVP comes at the cost of additional storage overhead compared to VP.
But as the effectiveness of ExtVP increases with smaller tables sizes (i.e.~higher selectivity), we can reduce this overhead to a large extent while retaining most of its benefits.
Let $SF$ be the \textit{selectivity factor} of a table in ExtVP, i.e.~its relative size compared to the corresponding VP table: $SF(ExtVP_{p_1|p_2}) = |ExtVP_{p_1|p_2}|/|VP_{p_1}|$. For example, $ExtVP_{follows|likes}^{OS}$ in Fig.~\ref{fig:extvp_all} has a $SF$ value of $0.25$ as its size is only a quarter of $VP_{follows}$.
Let $k = |\mathcal{P}|$ be the number of predicates in an RDF graph $G$ and $n = |G|$ be the number of triples in $G$. It holds that the sum of all tuples in VP, $|VP[G]|$, is also equal to $n$.
W.l.o.g.~assume all VP tables have equal size $n/k$ and $SF = 0.5$ for all ExtVP tables.
The size of an ExtVP schema for $G$ (i.e.~sum of all tuples) can be estimated:

\begin{tabbing}
$|ExtVP[G]|$ \= $= \underbrace{k}_{\text{\#predicates}} *~ (\underbrace{(3k-1)}_{\substack{\text{\#tables} \\ \text{per predicate}}} * \underbrace{\frac{n}{2k}}_{\text{table size}})$\\
\> $= (3k-1) * \frac{n}{2} < \frac{3}{2}kn$
\end{tabbing}

However, this is by far an overestimation of the real size as it assumes that all predicates can be combined with one another. In our experiments, typically more than $90 \%$ of all ExtVP tables were either empty or equal to VP and hence not stored. In general, the more predicates exist in an RDF dataset the more ExtVP tables will be empty as many of these predicates have distinct domains  (e.g.~predicates describing products vs.~users).
Exemplary, for a dataset with $n \approx 10^9$ triples and $86$ predicates the actual size of ExtVP was $\sim 11n$ (cf.~Sec.~\ref{sec:Evaluation}). HDFS storage space is normally not a limiting factor in an Hadoop environment and as we use the \textit{Parquet} columnar storage format in combination with \textit{snappy} compression to materialize the tables in HDFS, the physical size of ExtVP (including VP tables) was $\sim 1.3$ times the original input RDF dataset size in N-triples format.

Nonetheless, $11n$ tuples in ExtVP compared to $n$ tuples in VP states a significant overhead. On the one hand, tables with $SF \sim 1$ impose a large overhead while contributing only a negligible performance benefit. On the other hand, tables with $SF < 0.25$ give the best performance benefit while causing only little overhead. To this end, S2RDF supports the definition of a \textit{threshold} for $SF$ such that all ExtVP tables above this threshold are not considered. As demonstrated in Sec.~\ref{subsec:SFUpperBound}, a threshold of $0.25$ reduces the size of ExtVP from $\sim 11n$ to $\sim 2n$ tuples and at the same time already provides $95 \%$ of the performance benefit on average compared to using no threshold.

%% file: sections/querying.tex
Query processing in S2RDF is based on the algebraic representation of SPARQL expressions as defined in the official W3C recommendation~\cite{sparql_10}. We use \textit{Jena ARQ} to parse a SPARQL query into the corresponding algebra tree and apply some basic algebraic optimizations, e.g.~filter pushing. However, SPARQL query optimization was not a core aspect when developing S2RDF, hence there is still much room for improvement in this field. Finally, the tree is traversed from bottom up to generate the equivalent Spark SQL expressions based on our ExtVP schema described in Sec.~\ref{sec:ExtVP}.

\subsection{Basic Approach}
\label{subsec:QueryingBasicApproach}

\begin{figure*}[htb]
	\centering
	\includegraphics[width=1.0\textwidth]{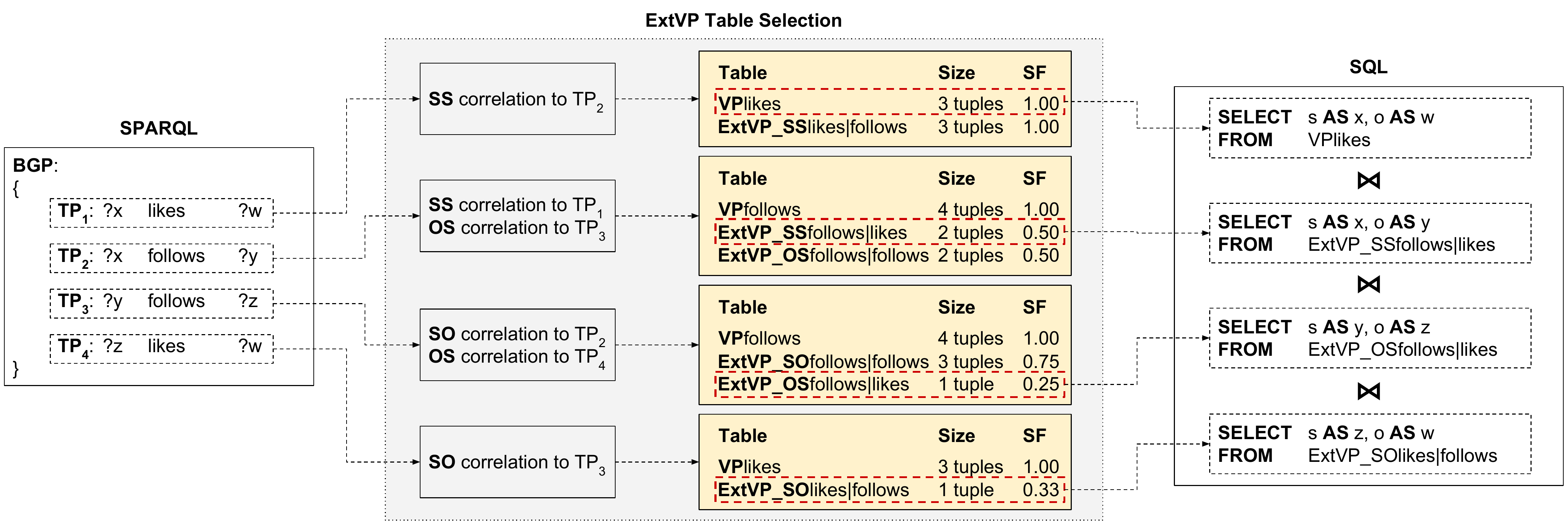}
	\caption{SPARQL to SQL for $Q_1$ based on ExtVP. Correlations between triple patterns determine the possible ExtVP tables. From these candidate tables, the ones with best (min) SF values get selected.}
	\label{fig:extvp_sql}
\end{figure*}

The basic procedure is that every triple pattern in a BGP $bgp = \{ tp_1, \ldots ,tp_n \}$ is represented by an equivalent subquery $\{ sq_1, \ldots ,sq_n \}$ and the results of these subqueries are joined to compute the result of $bgp$, $\Omega_{bgp} = sq_1 \Join \ldots \Join sq_n$.
For this, the query compiler of S2RDF has to select the appropriate table for every triple pattern $tp_i$. In a VP schema, this choice is unambiguous as it is simply defined by the predicate of $tp_i$ (cf.~Fig.~\ref{fig:vp_sql}). In an ExtVP schema, however, there are potentially several candidate tables defined by the correlations of $tp_i$ to other triple patterns in $bgp$. From these candidates, the table with the best selectivity factor $SF$ should be chosen.
S2RDF collects statistics about all tables in ExtVP during the initial creation process, most notably the selectivities ($SF$ values) and actual sizes (number of tuples), such that these statistics can be used for query generation. It also stores statistics about empty tables (which do not physically exist) as this empowers the query compiler to know that a query has no results without actually running it.

The table selection procedure is depicted in Algorithm~\ref{alg:TableSelection}. If the predicate of the input triple pattern $tp_i$ is a variable, we have to use the base triples table (or union of VP tables) to answer that pattern. If not, it initially starts with the corresponding VP table and iterates over all other triple patterns in the input BGP to check whether they have any correlation (SS, SO, OS) to $tp_i$. If $tp_i$ has more than one correlation to another triple pattern, the algorithm selects the corresponding ExtVP table with smallest (best) $SF$ value.

\begin{algorithm}[htb]
	\small
	\DontPrintSemicolon
	\SetAlgoNoEnd
	\SetKwInOut{Input}{input}
	\Input{ $TriplePattern~tp_i$ : $(s,p,o)$\\
	~$BGP$ : $Set \langle TriplePattern : (s,p,o) \rangle$}
	\SetKwInOut{Output}{output}
	\Output{ $tab$ : $Table$}
	\smallskip
	\lIf{$isVar(tp_i.p)$}{ \Return{$TriplesTable$} }\;
	\lElse{
		$tab \leftarrow VP_{tp_i.p}$ \tcp*[f]{initially start with VP table}
	}\;
	\smallskip
	\ForEach{$tp : TriplePattern \in BGP \neq tp_i$}{
		\If{$tp_i.s = tp.s \wedge SF(ExtVP_{tp_i.p|tp.p}^{SS}) < SF(tab)$}{
			$tab \leftarrow ExtVP_{tp_i.p|tp.p}^{SS}$ \tcp*[f]{SS correlation}
		}
		\If{$tp_i.s = tp.o \wedge SF(ExtVP_{tp_i.p|tp.p}^{SO}) < SF(tab)$}{
			$tab \leftarrow ExtVP_{tp_i.p|tp.p}^{SO}$ \tcp*[f]{SO correlation}
		}
		\If{$tp_i.o = tp.s \wedge SF(ExtVP_{tp_i.p|tp.p}^{OS}) < SF(tab)$}{
			$tab \leftarrow ExtVP_{tp_i.p|tp.p}^{OS}$ \tcp*[f]{OS correlation}
		}
	}
	\Return{$tab$}
	\caption{\textsc{TableSelection}}
	\label{alg:TableSelection}	
\end{algorithm}

For example, consider triple pattern $tp_3 = (?y, follows, ?z)$ in Fig.~\ref{fig:extvp_sql}. It has a SO correlation to $tp_2 = (?x, follows, ?y)$ on variable $?y$ and an OS correlation to $tp_4 = (?z, likes, ?w)$ on variable $?z$. Hence, in total, there are three candidate tables to answer $tp_3$: (1) $VP_{follows}$, (2) $ExtVP_{follows|follows}^{SO}$ and (3) $ExtVP_{follows|likes}^{OS}$.
From these tables, (3) gets selected as it has the best $SF$ value.

Once the appropriate table is selected, the corresponding SQL subquery to retrieve the results for triple pattern $tp_i$ can be derived from the position of variables and bound values in $tp_i$.
Bound values are used as conditions in the WHERE clause and variable names are used to rename table columns in the SELECT clause such that all subqueries can be easily joined on same column names, i.e.~using natural joins. The mapping of a triple pattern to SQL is depicted in Algorithm~\ref{alg:TP2SQL} using relational algebra notation. It checks all positions of $tp_i$ whether they contain a variable or bound value. In case of a variable, the corresponding column gets renamed by the variable name and added to a list of projections (we combine rename and projection for shorthand notation). For a bound value on subject or object position, the corresponding selection is added to a list of conditions. A bound predicate is already covered by the selected table and thus no additional selection is needed. In Fig.~\ref{fig:extvp_sql} the corresponding SQL subqueries for triple patterns $tp_1, \ldots, tp_4$ are given on the right.

\begin{algorithm}[htb]
	\small
	\DontPrintSemicolon
	\SetAlgoNoEnd
	\SetKwInOut{Input}{input}
	\Input{ $TriplePattern~tp_i$ : $(s,p,o)$\\
	~$tab$ : $Table$}
	\SetKwInOut{Output}{output}
	\Output{ $query$ : SQL (in relational algebra notation)}
	\smallskip
	$projections \leftarrow \emptyset$, $conditions \leftarrow \emptyset$\;
	\If{$isVar(tp_i.s)$}{
		$projections \leftarrow projections \cup (s \rightarrow tp_i.s)$
	}
	\lElse{ $conditions \leftarrow conditions \cup (s = tp_i.s)$ }\;
	\If{$isVar(tp_i.p)$}{
		$projections \leftarrow projections \cup (p \rightarrow tp_i.p)$
	}
	\If{$isVar(tp_i.o)$}{
		$projections \leftarrow projections \cup (o \rightarrow tp_i.o)$
	}
	\lElse{ $conditions \leftarrow conditions \cup (o = tp_i.o)$ }\;
	\Return{$query \leftarrow \pi[projections]\sigma[conditions](tab)$}
	\caption{\textsc{TP2SQL}}
	\label{alg:TP2SQL}	
\end{algorithm}

Finally, the subqueries for all triple patterns are joined to compute the BGP result. The overall procedure is depicted in Algorithm~\ref{alg:BGP2SQL}. For every triple pattern in a BGP it calls the table selection algorithm, maps it to the corresponding subquery using the selected table and joins it with the subqueries of the other triple patterns. If one of the selected tables is empty, i.e.~$SF = 0$, it can directly return an empty result without having to run the query. That means, a SPARQL query which contains a correlation between two predicates that does not exist in the dataset, can be answered by using the statistics only. 

\begin{algorithm}[htb]
	\small
	\DontPrintSemicolon
	\SetAlgoNoEnd
	\SetKwInOut{Input}{input}
	\Input{ $BGP$ : $Set \langle TriplePattern : (s,p,o) \rangle$ }
	\SetKwInOut{Output}{output}
	\Output{ $query$ : SQL (in relational algebra notation)}
	\smallskip
	$vars \leftarrow \emptyset$, $query \leftarrow \emptyset$\;
	\ForEach{$tp : TriplePattern \in BGP$}{
		$tab \leftarrow \textsc{TableSelection}(tp, BGP)$\;
		\lIf{$SF(tab) = 0$}{
			\Return{$\emptyset$}
		}\;
		\lElseIf{$vars = \emptyset$}{
			$query \leftarrow \textsc{TP2SQL}(tp, tab)$
		}\;
		\lElse{
			$query \leftarrow query \Join \textsc{TP2SQL}(tp, tab)$
		}\;
		$vars \leftarrow vars \cup vars(tp)$\;
	}
	\Return{$query$}
	\caption{\textsc{BGP2SQL}}
	\label{alg:BGP2SQL}	
\end{algorithm}

The remaining SPARQL 1.0 operators can be more or less directly mapped to the appropriate counterparts in Spark SQL. A \textsc{Filter} expression in SPARQL can be mapped to equivalent conditions in Spark SQL where we essentially have to adapt the SPARQL syntax to the syntax of SQL. These conditions can then be added to the WHERE clause of the corresponding (sub)query. \textsc{Optional} is realized by a left outer join while \textsc{Union}, \textsc{Offset}, \textsc{Limit}, \textsc{Order By} and \textsc{Distinct} are realized using their equivalent clauses in the SQL dialect of Spark.
S2RDF does currently not support the additional features introduced in SPARQL 1.1, e.g.~subqueries and aggregations. This is left for future work.

\subsection{Join Order Optimization}
\label{subsec:JoinOrderOptimization}

Regarding query semantics the order of triple patterns in a SPARQL BGP does not affect the query result, i.e.~a user can specify the triple patterns in any order. However, for query evaluation the order in which triple patterns are actually executed can have severe impacts on performance. For example, the join between two triple patterns that have no common variable results in a cross join.
Obviously, a query processor should avoid such execution plans, if possible. In S2RDF, for a BGP with $n$ triple patterns we need $n-1$ joins between ExtVP tables to compute the result. As query workload is typically I/O bound, it is crucial for efficiency reasons to reduce the amount of intermediate results in this join sequence. In fact, the BGP to SQL mapping depicted in Algorithm~\ref{alg:BGP2SQL} ignores the order of triple patterns and hence is likely to produce suboptimal query plans.

\begin{figure}[htb]
	\centering
	\includegraphics[width=1.0\columnwidth]{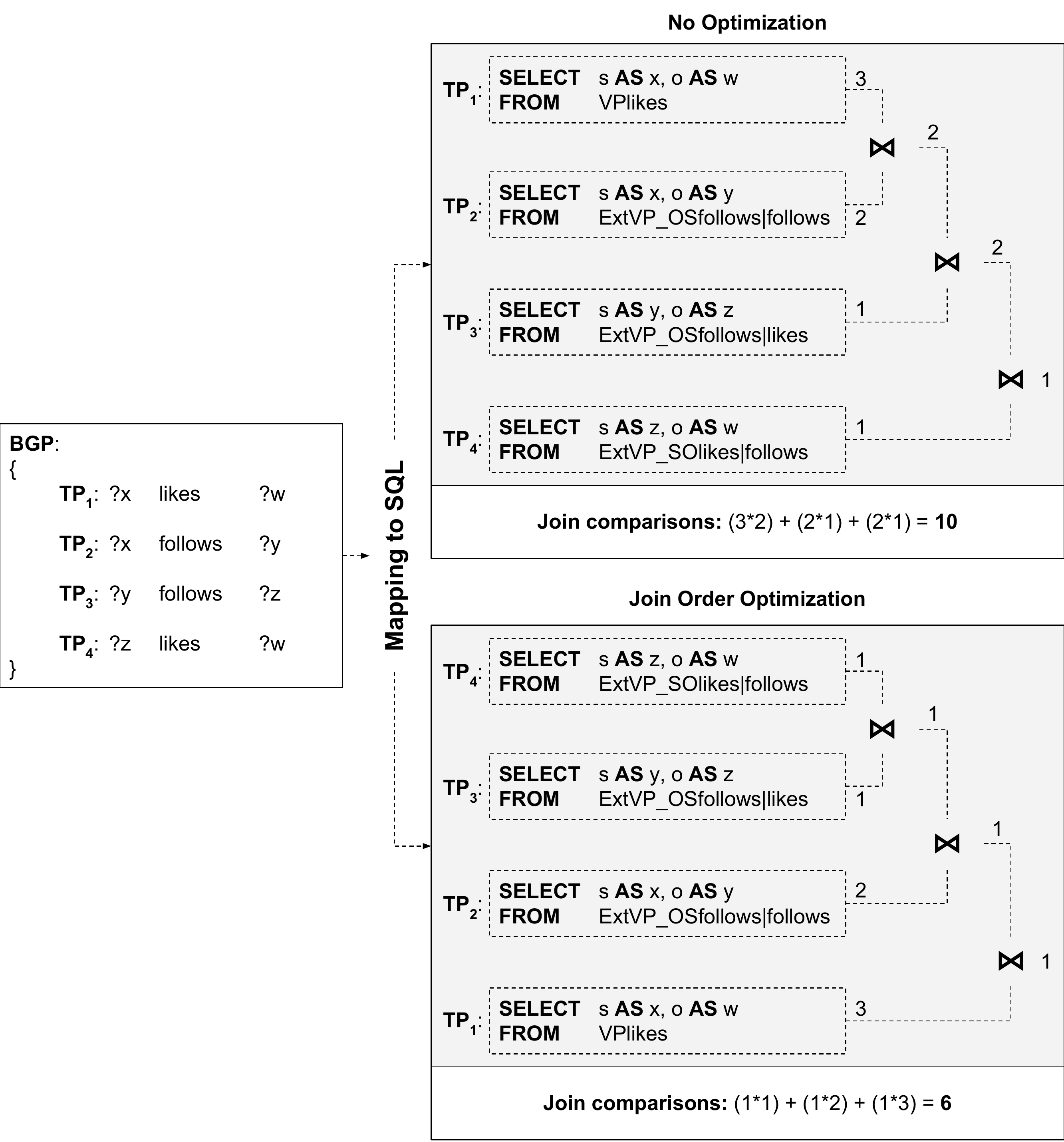}
	\caption{Join Order Optimization for query $Q_1$}
	\label{fig:join_order_opt}
\end{figure}

Let $sel(tp) = |\Omega_{tp}|/|G|$ be the \textit{selectivity} of a triple pattern $tp$ for RDF graph $G$. The general rule of thumb is to order triple patterns by selectivity, i.e.~$sel(tp_i) < sel(tp_{i+1})$. This is based on the assumption that smaller join inputs usually lead to smaller join outputs which is not always true but tend to be a good approximation in practice. Hence, the first and most obvious optimization is to make sure that patterns with more bound values are executed first and cross joins are avoided as they have the worst selectivity, i.e.~$\bigcup\limits_{j \leq i} vars(tp_j) \cap vars(tp_{i+1}) \neq \emptyset$. This can be derived from the query structure itself without knowing anything about the actual dataset.

In addition, S2RDF can make use of table statistics collected during initial ExtVP creation. As it knows the size of each VP and ExtVP table, it can order those triple patterns with the same amount of bound values by size of the corresponding table that is selected by Algorithm~\ref{alg:TableSelection}. It uses the actual table sizes (number of tuples) instead of selectivity factors as we want to join the smallest tables first and not the ones with the highest reduction compared to VP. The optimized version of Algorithm~\ref{alg:BGP2SQL} is depicted in Algorithm~\ref{alg:BGP2SQLopt}.

\begin{algorithm}[htb]
	\small
	\DontPrintSemicolon
	\SetAlgoNoEnd
	\SetKwInOut{Input}{input}
	\Input{ $BGP$ : $Set \langle TriplePattern : (s,p,o) \rangle$ }
	\SetKwInOut{Output}{output}
	\Output{ $query$ : SQL (in relational algebra notation)}
	\smallskip
	$vars \leftarrow \emptyset$, $query \leftarrow \emptyset$\;
	$tmpBGP \leftarrow orderByBoundValues(BGP)$\;
	\While{$tmpBGP \neq \emptyset$}{
		$tp_{next} \leftarrow \emptyset$, $tab_{next} \leftarrow \emptyset$\;
		\ForEach{$tp : TriplePattern \in tmpBGP$}{
			$tab \leftarrow \textsc{TableSelection}(tp, BGP)$\;
			\If{$tab_{next} = \emptyset~\vee$\\
			    $~~(Size(tab) < Size(tab_{next})~\wedge~vars \cap vars(tp) \neq \emptyset)$}{
				$tp_{next} \leftarrow tp$, $tab_{next} \leftarrow tab$
			}
		}
		\lIf{$SF(tab_{next}) = 0$}{
			\Return{$\emptyset$}
		}\;
		\ElseIf{$vars = \emptyset$}{
			$query \leftarrow \textsc{TP2SQL}(tp_{next}, tab_{next})$
		}
		\lElse{
			$query \leftarrow query \Join \textsc{TP2SQL}(tp_{next}, tab_{next})$
		}\;
		$vars \leftarrow vars \cup vars(tp_{next})$\;
		$tmpBGP \leftarrow tmpBGP \setminus \{tp_{next}\}$\;
	}
	\Return{$query$}
	\caption{\textsc{BGP2SQL\_opt}}
	\label{alg:BGP2SQLopt}	
\end{algorithm}

Consider again our running example in Fig.~\ref{fig:join_order_opt}. All triple patterns have the same amount of bound values and no cross join occurs when processing them in the order as listed. However, evaluating the patterns in given order would first join the two largest tables and produce two intermediate results. One of these results gets discarded in the very last step where it does not find a join partner.
As S2RDF knows the size of all tables, it instead joins the two smallest tables (corresponding to $tp_3$ and $tp_4$) first. Overall, this reduces the amount of intermediate results thus saving I/O, and also the total number of join comparisons thus saving CPU.

%% file: sections/evaluation.tex
The evaluation of S2RDF was performed on a small cluster of 10 machines (1 master and 9 worker), each equipped with a six core Intel Xeon E5-2420 CPU @1.90GHz, 2x2 TB disks, 32 GB RAM running Ubuntu 14.04 LTS and connected in a Gigabit Ethernet network. We used the Hadoop distribution of Cloudera CDH 5.4 including Spark 1.3. Each Spark executor was assigned 20 GB of memory, broadcast joins were disabled and the Parquet filter pushdown optimization enabled.

We compare our system with other state of the art distributed SPARQL query processors for Hadoop.
SHARD~\cite{rohloff_SHARD_2011} and PigSPARQL~\cite{pigsparql_2013} are based on MapReduce, Sempala~\cite{sempala_2014} uses Impala for query execution and H2RDF+~\cite{papailiou_H2RDF+_2013} is based on HBase.
In addition, we also compare S2RDF to a state of the art centralized RDF store, namely Virtuoso Open Source Edition v7.1.1~\cite{erling_virtuoso_2010} installed on single server with a four core Intel Xeon X5667 CPU @3.07GHz, 12 TB disk in hardware RAID 5 optimized for read performance and 32 GB RAM running Ubuntu 14.04 LTS.

The experiments were conducted on four datasets ranging from one million up to a billion RDF triples generated using the WatDiv data generator with scale factors 10, 100, 1000 and 10000, respectively. The generator is provided by the \textit{Waterloo SPARQL Diversity Test Suite}~\cite{watdiv_2014}, a more balanced stress testing environment for RDF data management systems with more diverse underlying workloads than other benchmarks. It covers all different query shapes (cf.~Sec.~\ref{subsec:RDF_SPARQL}) which allows us to test the performance of S2RDF and competitors more fine-grained.

The load times and store sizes are listed in Table~\ref{tab:datasets}.
For the largest dataset (SF10000), ExtVP consists of 2043 tables with 0<$SF$<1. The other potential tables were either empty (19780 with $SF=0$) or equal to their VP table (279 with $SF=1$) and hence not stored.
S2RDF needs significantly more time to load the data (using ExtVP) than the other systems due to the many semi-join operations that are performed. However, this is a one-time task and we have not spent much effort to optimize this process. In a production environment, one could think of a pay as you go approach where ExtVP tables are computed lazily on the fly when required for the first time in a query and materialized for usage in later queries. This way, there would be no initial loading overhead at the cost of performance slowdown for warm-up queries until the system converges.

\begin{table}[htb]
	\caption{WatDiv load times and HDFS sizes for S2RDF VP/ExtVP (0<SF<1) and competitors}
	\label{tab:datasets}
	\scriptsize
	\begin{tabularx}{\columnwidth}{llXXXX}
		\toprule
		& & \textbf{SF10} & \textbf{SF100} & \textbf{SF1000} & \textbf{SF10000} \\\midrule
		\multirow{3}{0mm}{\rotatebox{90}{\textbf{tuples}}} 
		&\textbf{original} & 1.08 M & 10.91 M & 109.2 M & 1091.5 M \\
		&\textbf{VP} & 1.08 M & 10.91 M & 109.2 M & 1091.5 M \\
		&\textbf{ExtVP} & 11.87 M & 119.94 M & 1197.9 M & 11967 M \\\midrule
		\multirow{7}{0mm}{\rotatebox{90}{\textbf{HDFS size}}} 
		&\textbf{original} & 49 MB & 507 MB & 5.3 GB & 54.9 GB \\
		&\textbf{VP} & 17 MB & 82 MB & 0.6 GB & 6.6 GB \\
		&\textbf{ExtVP} & 231 MB & 914 MB & 6.2 GB & 63.7 GB \\
		&\textbf{H2RDF+} & 47 MB & 517 MB & 5.2 GB & 57.0 GB \\
		&\textbf{Sempala} & 13 MB & 249 MB & 3.5 GB & 40.4 GB \\
		&\textbf{PigSPARQL} & 83 MB & 871 MB & 8.9 GB & 92.5 GB \\
		&\textbf{SHARD} & 96 MB & 981 MB & 9.9 GB & 100 GB \\\midrule
		\multirow{6}{0mm}{\rotatebox{90}{\textbf{load}}} 
		&\textbf{VP} & 72 s & 147 s & 290 s & 1065 s \\
		&\textbf{ExtVP} & 1430 s & 2418 s & 9497 s & 60572 s \\
		&\textbf{H2RDF+} & 122 s & 301 s & 507 s & 5425 s \\
		&\textbf{Sempala} & 26 s & 56 s & 333 s & 2782 s \\
		&\textbf{PigSPARQL} & 13 s & 16 s & 71 s & 498 s \\
		&\textbf{SHARD} & 19 s & 25 s & 134 s & 1222 s \\\midrule
		\multirow{3}{0mm}{\rotatebox{90}{\textbf{tables}}} 
		&\textbf{VP} & 86 & 86 & 86 & 86 \\
		&\textbf{ExtVP} & 2006 & 2038 & 2041 & 2043 \\
		&\textbf{total} & 2092 & 2124 & 2127 & 2129 \\\bottomrule
	\end{tabularx}
\end{table}

The outline of the evaluation is as follows:
We start with a performance comparison between VP and ExtVP in S2RDF to verify the benefits of ExtVP in Sec.~\ref{subsec:WatDivSelectivity}.
In Sec.~\ref{subsec:WatDivBasic} we present our results for the predefined WatDiv \textit{Basic Testing} use case in comparison to the other systems.
As large diameter queries are not adequately covered by the Basic Testing use case, we have designed an additional use case for WatDiv called \textit{Incremental Linear Testing} and present the results in Sec.~\ref{subsec:WatDivIncremental}.
Finally, in Sec.~\ref{subsec:SFUpperBound} we examine the effects of a selectivity threshold for ExtVP on size and performance and demonstrate that we can reduce the size of ExtVP to a large extent while retaining most of its performance.

\subsection{WatDiv Selectivity Testing (ST)}
\label{subsec:WatDivSelectivity}

First, we wanted to quantify the performance benefits of ExtVP in comparison to VP in S2RDF.
For this purpose, we have designed a set of carefully chosen queries to test the different characteristics of ExtVP, listed in Appendix~\ref{sec:AppendixWatDivSelectivity}.
Fig.~\ref{fig:wd10000_selectivity_coords} compares the runtimes of ExtVP and VP, the numbers are listed in Table~\ref{tab:wd_selectivity}.

\begin{figure}[htb]
	\centering
	\includegraphics[width=1.0\columnwidth]{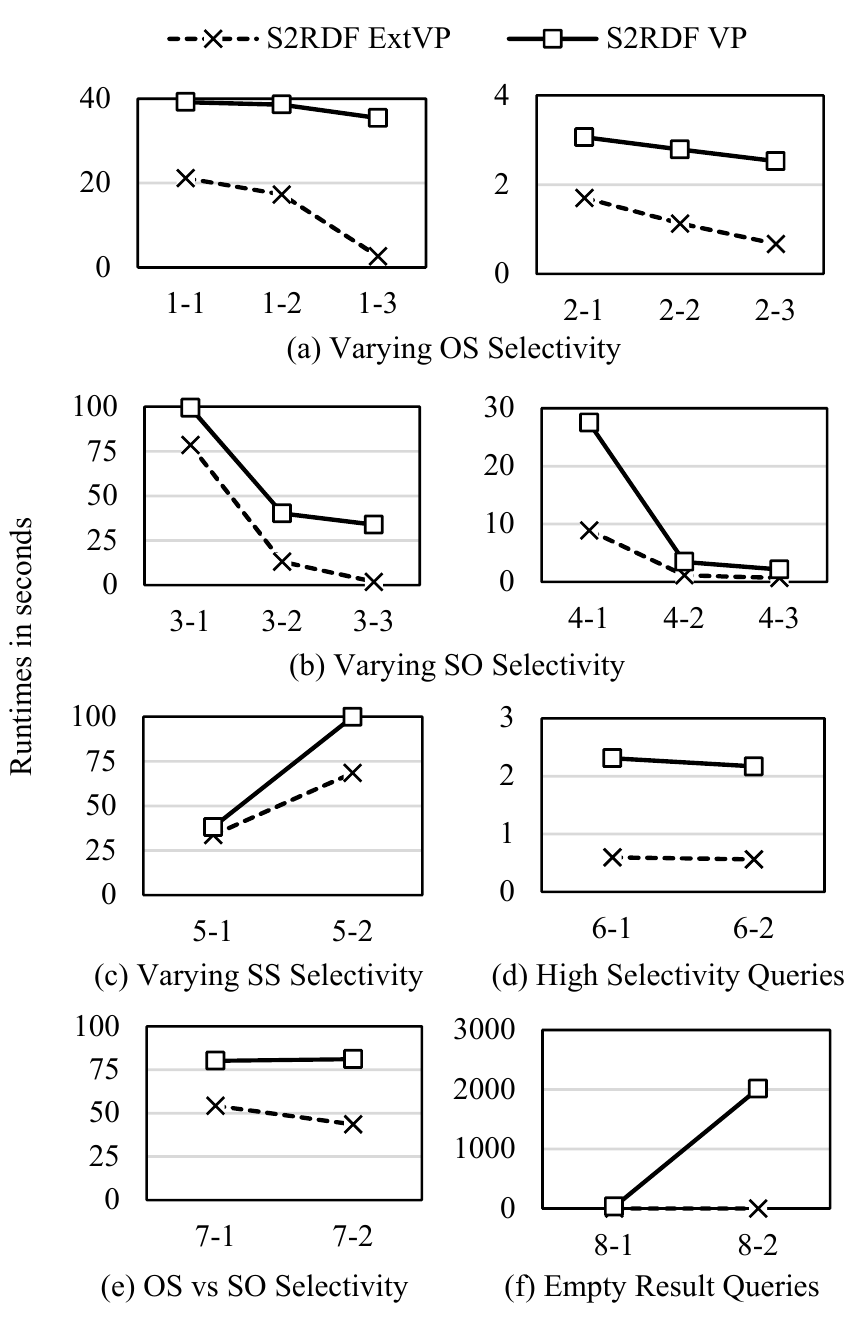}
	\caption{WatDiv Selectivity Testing (SF10000)}
	\label{fig:wd10000_selectivity_coords}
\end{figure}

\begin{table}[htb]
	\centering
	\scriptsize
	\caption{WatDiv Selectivity Testing comparing ExtVP with VP in S2RDF (runtimes in ms)}
	\setlength{\tabcolsep}{.05cm}
	\input{experiments/plots/Sel}
	\label{tab:wd_selectivity}
\end{table}

Queries ST-1-1 to ST-2-3 are used to verify the effectiveness of OS tables in ExtVP (cf.~Fig.~\ref{fig:wd10000_selectivity_coords}(a)).
ST-1-1 to ST-1-3 have the structure \verb|(?v0 friendOf ?v1 . ?v1 p ?v2)| where different predicates are used for \verb|p|. Thus, the input size of VP is large ($|VP_{friendOf}|=0.41*|G|$) while the selectivity factor $SF$ of the corresponding OS tables varies depending on \verb|p| (ST-1-1: 0.9, ST-1-2: 0.5, ST-1-3: 0.05).
ST-2-1 to ST-2-3 have the same structure but we replace \verb|friendOf| by \verb|reviewer| in the first triple pattern which has a much smaller input size in VP ($|VP_{reviewer}|=0.01*|G|$).
We expect the benefits of ExtVP to increase with lower $SF$ values which is clearly underpinned by the runtimes. For example, ST-1-3 is $\sim14$ times faster on ExtVP than on VP.
Of course, the benefit depends on the input size of the query in VP thus ST-1-X profits more than ST-2-X. But still, ST-2-3 is $\sim4$ times faster on ExtVP.

Queries ST-3-1 to ST-4-3 are used to verify the effectiveness of SO tables in ExtVP (cf.~Fig.~\ref{fig:wd10000_selectivity_coords}(b)).
The structure of ST-3-1 to ST-3-3 is \verb|(?v0 p ?v1 . ?v1 friendOf ?v2)|. Again, the choice of predicate for \verb|p| affects the selectivity of corresponding SO tables (ST-3-1: 0.9, ST-3-2: 0.31, ST-3-3: 0.04).
For ST-4-1 to ST-4-3 we replace \verb|friendOf| by \verb|likes| which has a much smaller input size in VP ($|VP_{likes}|=0.01*|G|$).
The outcome is similar to the OS tests before with ExtVP being $\sim18$ times faster for ST-3-3 and $\sim3$ times faster for ST-4-3.

Next, we test the effectiveness of SS tables in ExtVP (cf.~Fig.~\ref{fig:wd10000_selectivity_coords}(c)) using queries ST-5-1 and ST-5-2.
They have the structure \verb|(?v0 friendOf ?v1 . ?v0 p ?v2)| such that choice of \verb|p| affects the selectivity of corresponding SS tables (ST-5-1: 0.9, ST-5-2: 0.77).
As expected, the benefit of ExtVP is proportional to the selectivity of its tables compared to VP: $0.9*$VP for ST-5-1 and $0.7*$VP for ST-5-2.

ST-6-1 and ST-6-2 test the performance of queries with high selectivities ($SF < 0.01$) on small inputs (cf.~Fig.~\ref{fig:wd10000_selectivity_coords}(d)). ST-6-1 is a linear query whereas ST-6-2 is a star query. For both shapes, ExtVP is $\sim4$ times faster than VP and achieves sub-second runtimes on a billion triples.

ST-7-1 and ST-7-2 are designed to demonstrate that some queries benefit more from OS than from SO tables and vice versa. The queries consist of three linear-shaped triple patterns
\verb|(?v0 p1 ?v1 . ?v1 p2 ?v2 . ?v2 p3 ?v3)| such that S2RDF can choose between an SO or an OS table to answer the second triple pattern. For ST-7-1 the selectivity of OS is higher than SO and for ST-7-2 the selectivity of SO is higher than OS.

The last two queries, ST-8-1 and ST-8-2, are designed to demonstrate the usage of statistics. Using ExtVP the system can derive that both queries will not return any results without having to run them as the corresponding ExtVP tables are empty ($SF=0$). Especially ST-8-2 profits a lot from statistics because the first two triple patterns produce a large set of intermediate results in VP which get discarded afterwards.

Overall, the workload clearly underpins the desired characteristics of ExtVP compared to the baseline VP schema in all categories. As expected, the relative performance benefit is proportional to the selectivity of ExtVP tables compared to their corresponding VP counterparts.

\subsection{WatDiv Basic Testing}
\label{subsec:WatDivBasic}

\begin{figure*}[htbp]
	\centering
	\includegraphics[width=1.0\textwidth]{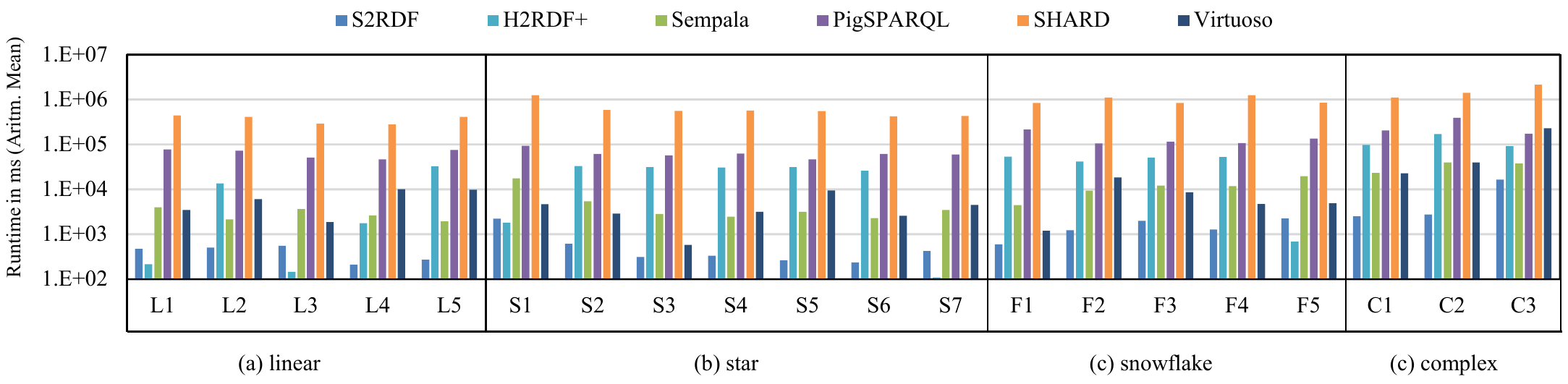}
	\caption{WatDiv Basic Testing (SF10000)}
	\label{fig:wd10000_basic}
\end{figure*}

\begin{table*}[htbp]
	\centering
	\scriptsize
	\caption{WatDiv Basic (in ms), AM-X = AM of category X, AM-T = total AM of all queries}
	\setlength{\tabcolsep}{.11cm}
	\input{experiments/plots/Basic-1}
	\newline
	\setlength{\tabcolsep}{.2cm}
	\input{experiments/plots/Basic-2}
	\label{tab:wd_basic}
\end{table*}

WatDiv comes with a set of 20 predefined query templates called \textit{Basic Testing} use case (listed in Appendix~\ref{sec:AppendixWatDivBasic}) which can be grouped in four categories according to their shape: \textit{star} (S), \textit{linear} (L), \textit{snowflake} (F) and \textit{complex} (C). We instantiated every query template several times for every dataset size and report the average mean (AM) runtime for all systems.
Fig.~\ref{fig:wd10000_basic} compares the different systems on the largest dataset (SF10000), corresponding AM runtimes are listed in Table~\ref{tab:wd_basic}.

We can observe that S2RDF outperforms PigSPARQL and SHARD by several orders of magnitude for all categories due to their underlying MapReduce-based batch execution engine. These systems scale very smooth with the data size but are not able to provide interactive query runtimes. PigSPARQL performs better than SHARD due to its multi-join optimization that can process several triple patterns in a single MapReduce job whereas SHARD uses one job per triple pattern.

The data model of Sempala is based on a unified property table and thus focuses on efficient star-shaped pattern execution where it needs no joins to retrieve the results.
Nonetheless, S2RDF outperforms Sempala by up to an order of magnitude for this category as well (S1-S7). We attribute this to the fact that Sempala, though not performing any joins, has to scan through the whole property table to find the matching rows thus query execution is limited by table scan time.
S2RDF, in contrast, performs a series of joins for star patterns but it can significantly reduce the input size as it can choose from a number of potential ExtVP SS tables and pick the one with the highest selectivity for every triple pattern.
For example, the star pattern of query S3 contains triple patterns with predicates \verb|rdf:type| and \verb|sorg:caption| and the corresponding table $ExtVP_{type|caption}^{SS}$ has selectivity $SF=0.02$, i.e.~it reduces the input size for triple pattern with predicate \verb|rdf:type| to only 2\%.
This underpins our finding from Sec.~\ref{sec:ExtVP} that input size reduction is often more effective than saving of join operations.
Sempala also shows a good performance for the other query types due to the underlying in-memory execution based on Impala but is not able to beat S2RDF for any query.
On average, S2RDF is an order of magnitude faster than Sempala for all four query categories.

The performance of H2RDF+ strongly depends on its execution model as some queries are executed centralized while others are executed using an optimized MapReduce job based on cost estimation.
For smaller datasets H2RDF+ can use efficient centralized merge joins and thus query runtimes are comparable to S2RDF.
However, with increasing data size many queries become too costly for centralized execution and MapReduce must be used instead.
For example, while both systems have nearly the same performance for query F1 on SF1000, S2RDF outperforms H2RDF+ by two orders of magnitude for the same query on SF10000.
The same is true for majority of queries regardless of their type.
There are only a few queries (5 out of 20) where runtimes of H2RDF+ are better than S2RDF on the largest dataset. These queries are highly selective returning only a few results and all of them contain a triple pattern where not only predicate but also subject or object are bound. Such queries are ideal candidates for centralized execution based on HBase lookups. But even for these queries, runtime differences of both systems are within the order of milliseconds.
On average, S2RDF outperforms H2RDF+ by at least one order of magnitude for all four query categories.

For completeness, we also executed the benchmark queries on centralized Virtuoso over the largest dataset (SF10000). We observed that runtimes of Virtuoso strongly correlate with the result size of a query and benefit from repeated execution.
Queries with empty results are generally very fast due to the use of sophisticated indexes.
Furthermore, if a query is executed several times for different template parameter values, its performance improves gradually probably due to caching effects. Thus, query runtimes of Virtuoso are rather skewed with relatively high standard deviations.
In contrast, the runtimes of S2RDF are pretty stable as they do not depend on the result size to such a large extent and we do not use query caching mechanisms. In Table~\ref{tab:wd_basic} we report both, the runtimes of Virtuoso for cold caches where we execute every query only once as well as average runtimes for repeated query execution.
Nonetheless, S2RDF is an order of magnitude faster than Virtuoso on average for all four query categories, even with repeated query execution.

Overall, the evaluation clearly demonstrates the superior performance of S2RDF compared to state of the art distributed and centralized RDF stores for all query shapes. In contrast to most existing approaches, ExtVP does not favor any specific query type and achieves consistent performance regardless of query pattern shape. Thus, S2RDF answers most of the benchmark queries in less than a second on a billion triples RDF graph.

\subsection{WatDiv Incremental Linear Testing (IL)}
\label{subsec:WatDivIncremental}

\begin{figure*}[htbp]
	\centering
	\includegraphics[width=1.0\textwidth]{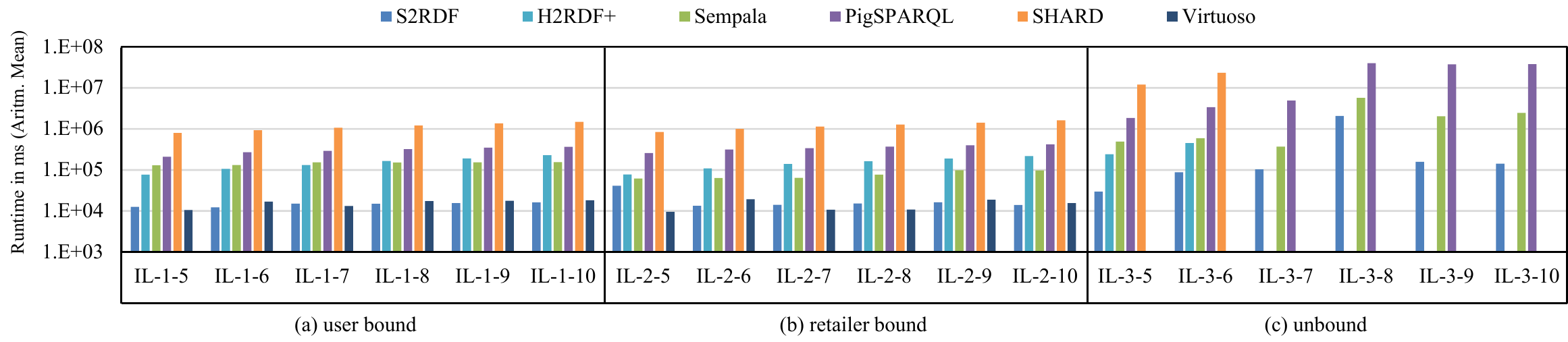}
	\caption{WatDiv Incremental Linear Testing (SF10000)}
	\label{fig:wd10000_incremental}
\end{figure*}

\begin{table*}[htbp]
	\centering
	\scriptsize
	\caption{WatDiv IL (in ms), AM-IL-X = AM of query type X, AM-Y = AM of queries with length Y}
	\setlength{\tabcolsep}{.05cm}
	\input{experiments/plots/Inc-1}
	\newline
	\setlength{\tabcolsep}{.045cm}
	\input{experiments/plots/Inc-2}
	\label{tab:wd_incremental}
\end{table*}

The queries of Basic Testing use case in WatDiv are well suited to test the performance for different query shapes but most of them have a rather small diameter. In fact, there are only two queries with a diameter larger than 3 (C1 and C2).
Most distributed RDF stores are optimized for small diameter queries as RDF datasets are typically fragmented based on hashing or graph partitioning algorithms that try to place vertices on the same cluster node that are close to each other in the RDF graph, e.g.~\cite{huang_HadoopRDF3X_2011, lee_SHAPE_2013, rohloff_SHARD_2011, sempala_2014, zhang_EAGRE_2013}.

Therefore, we wanted to have a more fine-grained look at the effects of query diameter on performance of S2RDF and competitors which was not possible with the given Basic Testing use case.
For that reason, we have designed an additional WatDiv use case called \textit{Incremental Linear Testing} focusing on linear queries with increasing size (diameter). It consists of three query types (IL-1, IL-2, IL-3) which are bound by user, retailer or unbound, resp. Each query type starts with a diameter of 5 (i.e.~5~triple patterns) and we incrementally add triple patterns (up to 10).
For example, query IL-1-8 is a user bound query with diameter 8.

All queries are listed in Appendix~\ref{sec:AppendixWatDivIncremental}. They were also added as an official use case to WatDiv and included in the WatDiv download package\footnote{\url{http://dsg.uwaterloo.ca/watdiv/}}. This emphasizes the importance of such a workload that was not adequately covered before.
Fig.~\ref{fig:wd10000_incremental} compares all systems on the largest dataset (SF10000), corresponding AM runtimes are listed in Table~\ref{tab:wd_incremental}.

The first query type (IL-1) starts from a given user following various edges along the graph, i.e.~it has the structure \verb|(u1 p1 ?v1 . ?v1 p2 ?v2 . ?v2 p3 ?v3 . ...)|.
We instantiate every query for 10 different users and we use the same users for all path lengths.
We can observe that S2RDF significantly outperforms all other distributed competitors while runtimes rise only slightly with increasing data size.
The queries make use of the two largest predicates in the dataset (\verb|friendOf| and \verb|follows|) that together represent ca. $70\%$ of all triples in the RDF graph ($0.7*|G|$).
This is probably also the reason why H2RDF+ does not use centralized execution for all queries starting from SF100.
By means of ExtVP, S2RDF can reduce the input size for predicate \verb|follows| from $0.3*|G|$ to only $0.07*|G|$ and from $0.4*|G|$ to $0.065*|G|$ for predicate \verb|friendOf| using a combination of OS and SO tables.
The only system that achieves similar runtimes as S2RDF is centralized Virtuoso with warm caches while being significantly slower for cold caches.

The second query type (IL-2) starts from a given retailer also following various edges along the graph, i.e.~it has the same structure as IL-1 but uses a different sequence of predicates along the path.
Again, we instantiate every query 10 times with different retailers.
Similar to IL-1, S2RDF clearly outperforms all other distributed competitors.
However, there is an interesting aspect when comparing the runtime of S2RDF for IL-2-5 with runtimes of IL-2-6 to IL-2-10. IL-2-5 performs much worse than all other queries despite being a subpattern of them. This seems to be odd at first glance but can be explained by looking at the query pattern.
IL-2-5 ends with following predicate (edge) \verb|friendOf| twice \verb|(... . ?v3 friendOf ?v4 . ?v4 friendOf ?v5)| which is the largest predicate in the graph ($0.4*|G|$).
Unfortunately, SO table $ExtVP_{friendOf|friendOf}^{SO}$ has selectivity $SF = 1$ which means that S2RDF cannot reduce the input size for the last triple pattern $t_5$ in the query and has to use table $VP_{friendOf}$ to retrieve the results for it.
IL-2-6 adds a triple pattern with predicate \verb|likes| to the end which now allows S2RDF to use OS table $ExtVP_{friendOf|likes}^{OS}$ with $SF = 0.24$ for triple pattern $t_5$ which reduces the input size for it from $0.4*|G|$ to $0.1*|G|$.
This example demonstrates that more triple patterns in a query can even lead to better performance in S2RDF despite the fact that it needs more joins.

The third query type (IL-3) has also the same structure as IL-1 and IL-2 but is not bound at all, i.e.~it starts from all vertices in the graph following various edges.
This type of query puts a heavy load on the system and produces very large result sets for large graphs.
Only S2RDF, Sempala and PigSPARQL were able to answer all queries up to diameter 10 on the largest dataset which demonstrate the excellent scalability of these systems.
Virtuoso was not able to answer any of the queries within a 10 hours timeout which confirms the scalability limitations of centralized RDF stores.
Interesting to see is that S2RDF runtime of IL-3-8 is an order of magnitude slower than for all other queries. There is a twofold reason for that.
First, the result size of IL-3-8 on SF10000 is extremely large ($\sim 25$ billion) and gets reduced by adding another triple pattern $t_9$ in IL-3-9 ($\sim 1$ billion).
Second, join order optimization of S2RDF (cf.~Sec.~\ref{subsec:JoinOrderOptimization}) does not simply add the join for $t_9$ in IL-3-9 to the end of join order but pushes it more to the beginning as its corresponding ExtVP table is rather small.
As a result, S2RDF avoids to compute the same output as for IL-3-8 and then joining with $t_9$ to compute the final result for IL-3-9 but instead reduces the intermediate result size significantly.

Overall, S2RDF demonstrates its superior performance and scalability for queries with large diameters, a query type that is not sufficiently covered by the Basic Testing use case and is also underrepresented in other existing RDF/SPARQL benchmarks.
Our new use case reveals that performance of most distributed RDF stores significantly drops for such workloads as their data model is optimized to answer small diameter queries.
It is now included in WatDiv and we hope that this will accelerate a better support in future.

\subsection{SF Threshold}
\label{subsec:SFUpperBound}

\begin{figure*}[htb]
	\centering
	\includegraphics[width=1.0\textwidth]{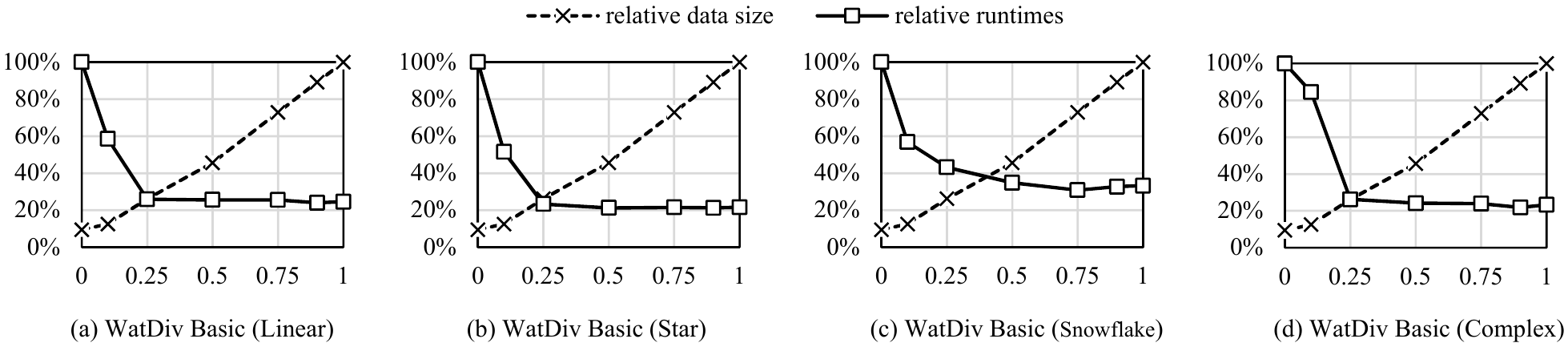}
	\caption{Relative runtimes and storage sizes in relation to increasing SF threshold values}
	\label{fig:sfub}
\end{figure*}

In this section we investigate the influence of a threshold for selectivity factor $SF$ of tables in ExtVP (SF TH) as discussed in Sec.~\ref{subsec:ExtVPSize} on runtime and storage size with S2RDF.
For example, a threshold of $0.5$ for $SF$ means that only those ExtVP tables with $SF < 0.5$ are materialized.

\begin{table}[htb]
	\centering
	\scriptsize
	\caption{ExtVP sizes (including VP) for varying SF threshold values (SF10000)}
	\begin{tabularx}{\columnwidth}{rrrr}
		\toprule
		\textbf{SF TH} & \textbf{\# tables (in \%)} & \textbf{\# tuples} & \textbf{HDFS size (in \%)} \\
		\midrule
		\textbf{0.00} & 86 ~(4\%) & 1091M & 6.6 GB (9\%) \\
		\textbf{0.10} & 924 (43\%) & 1508M & 8.8 GB (13\%) \\
		\textbf{0.25} & 1275 (60\%) & 3316M & 18.4 GB (26\%) \\
		\textbf{0.50} & 1609 (76\%) & 5889M & 32.1 GB (46\%) \\
		\textbf{0.75} & 1887 (89\%) & 9480M & 51.3 GB (73\%) \\
		\textbf{0.90} & 2006 (94\%) & 11631M & 62.8 GB (89\%) \\
		\textbf{1.00} & 2129 (100\%) & 13059M & 70.4 GB (100\%) \\
		\bottomrule
	\end{tabularx}%
	\label{tab:sfub}%
\end{table}%

Table~\ref{tab:sfub} gives an overview of the number of tables maintained by S2RDF (ExtVP + VP), the total number of table tuples and total storage size in HDFS.
SF TH = 0 means that no ExtVP tables are stored at all and hence only VP is used whereas SF TH = 1 means that all tables with $SF < 1$ are considered.
The best performance, of course, is achieved for SF TH = 1 but this also means that we store a lot of tables with low selectivities that cost a lot of storage but do not improve performance to a large extent.
We test the runtimes of S2RDF for WatDiv Basic Testing use case on SF10000 for various SF TH values.
A comparison of relative runtimes (grouped by query category) and actual data size with respect to SF TH values is illustrated in Fig.~\ref{fig:sfub}.

It turns out that for SF TH = 0.25 we already achieve most of the performance benefit compared to the baseline execution with SF TH = 0 (which corresponds to VP). In fact, on average we achieve $95\%$ of the performance benefit that we get when using no threshold. At the same time, we maintain only $3n$ tuples in total over all tables (including VP) with $n$ being the number of triples in the RDF graph, compared to $12n$ for SF TH = 1. Thus, ExtVP implies an overhead of only $2n$ compared to VP for SF TH = 0.25 which also directly corresponds to physical storage in HDFS.

If we compare the different query categories (cf.~Fig.~\ref{fig:sfub}) we can observe that (a) linear-, (b) star- and (d) complex-shaped queries hardly improve anymore for SF TH values larger than 0.25. Only the (c) snowflake-shaped queries noticeably profit from larger values, more precisely F3 and F5. Indeed, both contain several triple patterns where corresponding ExtVP tables have selectivity $0.25 < SF < 0.5$.

In summary, the performance of S2RDF when executed on ExtVP with a threshold of 0.25 for $SF$ is almost the same compared to ExtVP with no threshold on $SF$ while using only $25\%$ of table tuples and storage size.

%% file: experiments/plots/Sel.tex
\begin{tabularx}{\columnwidth}{lrr|rr|rr|rr}
\toprule
      & \multicolumn{2}{c|}{\textbf{SF10}} & \multicolumn{2}{c|}{\textbf{SF100}} & \multicolumn{2}{c|}{\textbf{SF1000}} & \multicolumn{2}{c}{\textbf{SF10000}} \\
\midrule
\textbf{Query} & \textbf{ExtVP} & \textbf{VP} & \textbf{ExtVP} & \textbf{VP} & \textbf{ExtVP} & \textbf{VP} & \textbf{ExtVP} & \textbf{VP} \\
\midrule
\textbf{ST-1-1} & 410   & 471   & 997   & 1126  & 3486  & 4233  & 21147 & 39179 \\
\textbf{ST-1-2} & 400   & 531   & 922   & 967   & 3008  & 4022  & 17266 & 38601 \\
\textbf{ST-1-3} & 403   & 432   & 353   & 889   & 623   & 3811  & 2616  & 35405 \\
\textbf{ST-2-1} & 133   & 143   & 295   & 202   & 584   & 762   & 1704  & 3063 \\
\textbf{ST-2-2} & 209   & 201   & 224   & 191   & 419   & 697   & 1129  & 2786 \\
\textbf{ST-2-3} & 183   & 127   & 203   & 204   & 300   & 633   & 672   & 2526 \\
\midrule
\textbf{ST-3-1} & 527   & 539   & 1368  & 1644  & 7498  & 8496  & 78525 & 99282 \\
\textbf{ST-3-2} & 278   & 371   & 383   & 926   & 1774  & 3632  & 13112 & 40106 \\
\textbf{ST-3-3} & 193   & 306   & 210   & 845   & 432   & 3152  & 1861  & 33831 \\
\textbf{ST-4-1} & 220   & 281   & 378   & 775   & 1200  & 3174  & 8911  & 27487 \\
\textbf{ST-4-2} & 124   & 127   & 197   & 196   & 312   & 674   & 1145  & 3455 \\
\textbf{ST-4-3} & 140   & 119   & 201   & 191   & 294   & 642   & 677   & 2142 \\
\midrule
\textbf{ST-5-1} & 346   & 313   & 818   & 801   & 3399  & 3594  & 33684 & 38138 \\
\textbf{ST-5-2} & 507   & 537   & 1283  & 1621  & 7298  & 7783  & 68443 & 99886 \\
\midrule
\textbf{ST-6-1} & 110   & 118   & 220   & 172   & 285   & 535   & 596   & 2311 \\
\textbf{ST-6-2} & 115   & 119   & 196   & 194   & 274   & 534   & 562   & 2169 \\
\midrule
\textbf{ST-7-1} & 392   & 575   & 922   & 1360  & 4479  & 7770  & 54259 & 80141 \\
\textbf{ST-7-2} & 381   & 465   & 861   & 1422  & 3919  & 6196  & 43507 & 81083 \\
\midrule
\textbf{ST-8-1} & 0     & 433   & 0     & 926   & 0     & 3484  & 0     & 31978 \\
\textbf{ST-8-2} & 0     & 2350  & 0     & 24205 & 0     & 92879 & 0     & 2012104 \\
\bottomrule
\end{tabularx}%

%% file: experiments/plots/Basic-1.tex
\begin{tabularx}{\textwidth}{rrrrrrr|r|rrrrrrr|r}
\toprule
      & \textbf{Query} & \textbf{L1} & \textbf{L2} & \textbf{L3} & \textbf{L4} & \textbf{L5} & \textbf{AM-L} & \textbf{S1} & \textbf{S2} & \textbf{S3} & \textbf{S4} & \textbf{S5} & \textbf{S6} & \textbf{S7} & \textbf{AM-S} \\
\midrule
\multicolumn{1}{c}{\multirow{6}[1]{*}{\begin{sideways}\textbf{SF10}\end{sideways}}} & \textbf{S2RDF ExtVP} & 164   & 145   & 97    & 95    & 140   & 128   & 478   & 204   & 180   & 190   & 211   & 138   & 141   & 220 \\
\multicolumn{1}{c}{} & \textbf{S2RDF VP} & 164   & 134   & 98    & 96    & 147   & 128   & 510   & 191   & 185   & 187   & 176   & 142   & 150   & 220 \\
\multicolumn{1}{c}{} & \textbf{H2RDF+} & 78    & 82    & 55    & 45    & 82    & 68    & 1448  & 73    & 70    & 86    & 59    & 55    & 49    & 263 \\
\multicolumn{1}{c}{} & \textbf{Sempala} & 786   & 748   & 642   & 642   & 758   & 715   & 1120  & 822   & 764   & 790   & 866   & 714   & 652   & 818 \\
\multicolumn{1}{c}{} & \textbf{PigSPARQL} & 68000 & 67500 & 41500 & 41500 & 67500 & 57200 & 42000 & 41000 & 41500 & 41000 & 41500 & 41500 & 41500 & 41429 \\
\multicolumn{1}{c}{} & \textbf{SHARD} & 77084 & 75243 & 56547 & 56051 & 76969 & 68379 & 183644 & 93484 & 87026 & 95513 & 88523 & 72573 & 70062 & 98689 \\
\midrule
\multicolumn{1}{c}{\multirow{6}[2]{*}{\begin{sideways}\textbf{SF100}\end{sideways}}} & \textbf{S2RDF ExtVP} & 168   & 193   & 126   & 107   & 173   & 153   & 562   & 216   & 214   & 221   & 193   & 146   & 164   & 245 \\
\multicolumn{1}{c}{} & \textbf{S2RDF VP} & 192   & 167   & 140   & 127   & 186   & 162   & 644   & 237   & 214   & 234   & 244   & 199   & 184   & 279 \\
\multicolumn{1}{c}{} & \textbf{H2RDF+} & 91    & 229   & 74    & 105   & 287   & 157   & 1121  & 394   & 273   & 275   & 124   & 110   & 50    & 335 \\
\multicolumn{1}{c}{} & \textbf{Sempala} & 750   & 748   & 636   & 646   & 748   & 706   & 1340  & 854   & 756   & 844   & 940   & 764   & 754   & 893 \\
\multicolumn{1}{c}{} & \textbf{PigSPARQL} & 68000 & 68000 & 41000 & 41500 & 67500 & 57200 & 41500 & 41500 & 41500 & 41000 & 41500 & 41000 & 41000 & 41286 \\
\multicolumn{1}{c}{} & \textbf{SHARD} & 79008 & 74564 & 58076 & 52523 & 76045 & 68043 & 185660 & 90096 & 87010 & 92665 & 87978 & 74571 & 70571 & 98364 \\
\midrule
\multicolumn{1}{c}{\multirow{6}[2]{*}{\begin{sideways}\textbf{SF1000}\end{sideways}}} & \textbf{S2RDF ExtVP} & 202   & 196   & 196   & 132   & 162   & 178   & 735   & 294   & 219   & 209   & 199   & 209   & 191   & 294 \\
\multicolumn{1}{c}{} & \textbf{S2RDF VP} & 455   & 284   & 339   & 167   & 318   & 313   & 1584  & 484   & 352   & 457   & 338   & 413   & 402   & 576 \\
\multicolumn{1}{c}{} & \textbf{H2RDF+} & 98    & 1450  & 62    & 207   & 1616  & 687   & 942   & 3831  & 2423  & 1937  & 996   & 673   & 51    & 1550 \\
\multicolumn{1}{c}{} & \textbf{Sempala} & 904   & 746   & 740   & 664   & 752   & 761   & 3130  & 1058  & 862   & 876   & 960   & 848   & 870   & 1229 \\
\multicolumn{1}{c}{} & \textbf{PigSPARQL} & 68000 & 67500 & 46500 & 41000 & 68000 & 58200 & 57000 & 46000 & 44000 & 46500 & 42000 & 46500 & 46500 & 46929 \\
\multicolumn{1}{c}{} & \textbf{SHARD} & 103153 & 99905 & 72603 & 67937 & 99595 & 88639 & 264976 & 130625 & 122958 & 127587 & 123832 & 99866 & 102040 & 138840 \\
\midrule
\multicolumn{1}{c}{\multirow{8}[1]{*}{\begin{sideways}\textbf{SF10000}\end{sideways}}} & \textbf{S2RDF ExtVP} & 471   & 498   & 549   & 209   & 270   & 399   & 2208  & 607   & 311   & 329   & 260   & 235   & 420   & 624 \\
\multicolumn{1}{c}{} & \textbf{S2RDF VP} & 1966  & 1391  & 1461  & 353   & 1372  & 1309  & 8873  & 1712  & 974   & 1232  & 760   & 1828  & 1856  & 2462 \\
\multicolumn{1}{c}{} & \textbf{H2RDF+} & 214   & 13447 & 145   & 1752  & 32222 & 9556  & 1809  & 32902 & 30844 & 30419 & 30807 & 25795 & 109   & 21812 \\
\multicolumn{1}{c}{} & \textbf{Sempala} & 3938  & 2140  & 3630  & 2616  & 1914  & 2848  & 17386 & 5368  & 2816  & 2442  & 3142  & 2260  & 3476  & 5270 \\
\multicolumn{1}{c}{} & \textbf{PigSPARQL} & 76500 & 72500 & 51000 & 46500 & 75000 & 64300 & 92500 & 61000 & 56500 & 62000 & 46500 & 61500 & 59000 & 62714 \\
\multicolumn{1}{c}{} & \textbf{SHARD} & 444511 & 406007 & 288980 & 275534 & 406591 & 364324 & 1240267 & 579207 & 555911 & 567297 & 550959 & 424402 & 428905 & 620992 \\
\multicolumn{1}{c}{} & \textbf{Virtuoso (cold)} & 16009 & 15727 & 7580  & 46287 & 44260 & 25973 & 15873 & 13072 & 2726  & 11445 & 1910  & 6779  & 7084  & 8413 \\
\multicolumn{1}{c}{} & \textbf{Virtuoso (AM)} & 3437  & 5984  & 1863  & 10055 & 9710  & 6210  & 4630  & 2853  & 576   & 3157  & 9470  & 2555  & 4516  & 3965 \\
\\
\end{tabularx}%

%% file: experiments/plots/Basic-2.tex
\begin{tabularx}{\textwidth}{rrrrrrr|r|rrr|r|r}
\toprule
      & \textbf{Query} & \textbf{F1} & \textbf{F2} & \textbf{F3} & \textbf{F4} & \textbf{F5} & \textbf{AM-F} & \textbf{C1} & \textbf{C2} & \textbf{C3} & \textbf{AM-C} & \textbf{AM-T} \\
\midrule
\multicolumn{1}{c}{\multirow{6}[1]{*}{\begin{sideways}\textbf{SF10}\end{sideways}}} & \textbf{S2RDF ExtVP} & 370   & 451   & 376   & 461   & 334   & 398   & 535   & 472   & 450   & 486   & 282 \\
\multicolumn{1}{c}{} & \textbf{S2RDF VP} & 306   & 396   & 300   & 437   & 314   & 350   & 578   & 578   & 693   & 616   & 289 \\
\multicolumn{1}{c}{} & \textbf{H2RDF+} & 86    & 108   & 113   & 169   & 135   & 122   & 232   & 391   & 1386  & 670   & 240 \\
\multicolumn{1}{c}{} & \textbf{Sempala} & 866   & 1100  & 1180  & 1172  & 988   & 1061  & 1184  & 1354  & 1066  & 1201  & 911 \\
\multicolumn{1}{c}{} & \textbf{PigSPARQL} & 85000 & 80000 & 78500 & 80500 & 80500 & 80900 & 98500 & 119500 & 42000 & 86667 & 62025 \\
\multicolumn{1}{c}{} & \textbf{SHARD} & 134813 & 172248 & 133469 & 191114 & 122713 & 150871 & 168505 & 205405 & 126769 & 166893 & 114388 \\
\midrule
\multicolumn{1}{c}{\multirow{6}[2]{*}{\begin{sideways}\textbf{SF100}\end{sideways}}} & \textbf{S2RDF ExtVP} & 393   & 539   & 385   & 579   & 398   & 459   & 577   & 689   & 688   & 651   & 337 \\
\multicolumn{1}{c}{} & \textbf{S2RDF VP} & 388   & 478   & 443   & 604   & 425   & 468   & 582   & 758   & 1274  & 871   & 386 \\
\multicolumn{1}{c}{} & \textbf{H2RDF+} & 167   & 320   & 330   & 566   & 241   & 325   & 1180  & 2096  & 29280 & 10852 & 1866 \\
\multicolumn{1}{c}{} & \textbf{Sempala} & 928   & 1174  & 1150  & 1202  & 1072  & 1105  & 1292  & 1656  & 1702  & 1550  & 998 \\
\multicolumn{1}{c}{} & \textbf{PigSPARQL} & 89500 & 80000 & 80500 & 80000 & 80000 & 82000 & 97500 & 116500 & 51000 & 88333 & 62500 \\
\multicolumn{1}{c}{} & \textbf{SHARD} & 133293 & 173122 & 130420 & 184154 & 124463 & 149090 & 167754 & 206338 & 139868 & 171320 & 114409 \\
\midrule
\multicolumn{1}{c}{\multirow{6}[2]{*}{\begin{sideways}\textbf{SF1000}\end{sideways}}} & \textbf{S2RDF ExtVP} & 433   & 642   & 638   & 692   & 672   & 615   & 923   & 1460  & 2929  & 1771  & 567 \\
\multicolumn{1}{c}{} & \textbf{S2RDF VP} & 733   & 949   & 946   & 1068  & 1116  & 962   & 1338  & 2017  & 6372  & 3242  & 1007 \\
\multicolumn{1}{c}{} & \textbf{H2RDF+} & 476   & 2926  & 1952  & 5642  & 203   & 2240  & 53032 & 20250 & 35617 & 36300 & 6719 \\
\multicolumn{1}{c}{} & \textbf{Sempala} & 1068  & 1704  & 1538  & 1950  & 2545  & 1761  & 2828  & 5992  & 6040  & 4953  & 1804 \\
\multicolumn{1}{c}{} & \textbf{PigSPARQL} & 97500 & 80500 & 92500 & 85500 & 99000 & 91000 & 119500 & 272500 & 64000 & 152000 & 76525 \\
\multicolumn{1}{c}{} & \textbf{SHARD} & 182215 & 237370 & 181606 & 266426 & 181581 & 209839 & 236828 & 297344 & 298759 & 277644 & 164860 \\
\midrule
\multicolumn{1}{c}{\multirow{8}[1]{*}{\begin{sideways}\textbf{SF10000}\end{sideways}}} & \textbf{S2RDF ExtVP} & 590   & 1226  & 1969  & 1265  & 2254  & 1461  & 2508  & 2740  & 16407 & 7218  & 1766 \\
\multicolumn{1}{c}{} & \textbf{S2RDF VP} & 2696  & 2859  & 4815  & 3744  & 6271  & 4077  & 5194  & 9865  & 58416 & 24492 & 5882 \\
\multicolumn{1}{c}{} & \textbf{H2RDF+} & 52843 & 41480 & 51085 & 52323 & 679   & 39682 & 96777 & 170473 & 91189 & 119480 & 37866 \\
\multicolumn{1}{c}{} & \textbf{Sempala} & 4420  & 9316  & 12090 & 11668 & 19516 & 11402 & 23136 & 39710 & 37462 & 33436 & 10422 \\
\multicolumn{1}{c}{} & \textbf{PigSPARQL} & 212000 & 105500 & 115500 & 106000 & 133500 & 134500 & 205000 & 387000 & 172000 & 254667 & 109850 \\
\multicolumn{1}{c}{} & \textbf{SHARD} & 835514 & 1103997 & 834778 & 1245115 & 841380 & 972157 & 1100372 & 1398989 & 2146928 & 1548763 & 783782 \\
\multicolumn{1}{c}{} & \textbf{Virtuoso (cold)} & 3997  & 30627 & 31144 & 11374 & 11362 & 17701 & 22801 & 39711 & 226129 & 96214 & 28295 \\
\multicolumn{1}{c}{} & \textbf{Virtuoso (AM)} & 1197  & 18376 & 8629  & 4705 & 4879  & 7557  & 22801 & 39711 & 226129 & 96214 & 19262 \\
\bottomrule
\end{tabularx}%

%% file: experiments/plots/Inc-1.tex
\begin{tabularx}{\textwidth}{rrrrrrrr|r|rrrrrr|r}
 \toprule
      & \textbf{Query} & \textbf{IL-1-5} & \textbf{IL-1-6} & \textbf{IL-1-7} & \textbf{IL-1-8} & \textbf{IL-1-9} & \textbf{IL-1-10} & \textbf{AM-IL-1} & \textbf{IL-2-5} & \textbf{IL-2-6} & \textbf{IL-2-7} & \textbf{IL-2-8} & \textbf{IL-2-9} & \textbf{IL-2-10} & \textbf{AM-IL-2} \\
\midrule
\multicolumn{1}{c}{\multirow{6}[1]{*}{\begin{sideways}\textbf{SF10}\end{sideways}}} & \textbf{S2RDF ExtVP} & 307   & 360   & 390   & 493   & 503   & 701   & 459   & 453   & 334   & 388   & 466   & 495   & 660   & 466 \\
\multicolumn{1}{c}{} & \textbf{S2RDF VP} & 387   & 490   & 552   & 608   & 671   & 749   & 576   & 590   & 642   & 696   & 727   & 782   & 850   & 714 \\
\multicolumn{1}{c}{} & \textbf{H2RDF+} & 523   & 1549  & 1722  & 1585  & 1640  & 2032  & 1508  & 689   & 1949  & 1818  & 1901  & 1990  & 2090  & 1740 \\
\multicolumn{1}{c}{} & \textbf{Sempala} & 1123  & 1065  & 1174  & 1291  & 1361  & 1444  & 1243  & 1054  & 1050  & 1062  & 1079  & 1081  & 1191  & 1086 \\
\multicolumn{1}{c}{} & \textbf{PigSPARQL} & 118800 & 149122 & 171543 & 196654 & 219919 & 243330 & 183228 & 128911 & 161337 & 184990 & 214918 & 244303 & 272150 & 201101 \\
\multicolumn{1}{c}{} & \textbf{SHARD} & 114427 & 136318 & 152320 & 173558 & 189693 & 209043 & 162560 & 112360 & 129302 & 145754 & 168682 & 184774 & 208121 & 158165 \\
\midrule
\multicolumn{1}{c}{\multirow{6}[2]{*}{\begin{sideways}\textbf{SF100}\end{sideways}}} & \textbf{S2RDF ExtVP} & 558   & 604   & 711   & 834   & 875   & 1299  & 814   & 969   & 594   & 654   & 795   & 911   & 1047  & 828 \\
\multicolumn{1}{c}{} & \textbf{S2RDF VP} & 942   & 1179  & 1291  & 1347  & 1486  & 1596  & 1307  & 1451  & 1655  & 1735  & 1864  & 2000  & 2090  & 1799 \\
\multicolumn{1}{c}{} & \textbf{H2RDF+} & 23498 & 27949 & 36826 & 28524 & 46006 & 29446 & 32041 & 23463 & 33042 & 44801 & 34200 & 71949 & 35621 & 40513 \\
\multicolumn{1}{c}{} & \textbf{Sempala} & 3643  & 3753  & 3844  & 3919  & 4042  & 4126  & 3888  & 2164  & 2257  & 2290  & 2450  & 2527  & 2644  & 2389 \\
\multicolumn{1}{c}{} & \textbf{PigSPARQL} & 128666 & 153755 & 178091 & 205638 & 228214 & 249651 & 190669 & 167104 & 209958 & 230734 & 254635 & 281279 & 304153 & 241310 \\
\multicolumn{1}{c}{} & \textbf{SHARD} & 118494 & 136505 & 154412 & 173364 & 192923 & 213625 & 164887 & 117176 & 130308 & 148463 & 170810 & 184851 & 208468 & 160013 \\
\midrule
\multicolumn{1}{c}{\multirow{6}[2]{*}{\begin{sideways}\textbf{SF1000}\end{sideways}}} & \textbf{S2RDF ExtVP} & 1724  & 1745  & 1965  & 2029  & 2185  & 2643  & 2048  & 4944  & 1869  & 1980  & 2114  & 2382  & 2413  & 2617 \\
\multicolumn{1}{c}{} & \textbf{S2RDF VP} & 4517  & 6108  & 6376  & 6679  & 7214  & 7045  & 6323  & 8006  & 9250  & 9691  & 9981  & 10414 & 10070 & 9568 \\
\multicolumn{1}{c}{} & \textbf{H2RDF+} & 25351 & 48957 & 78117 & 71448 & 99636 & 92232 & 69290 & 25090 & 49745 & 50514 & 73750 & 122180 & 99416 & 70116 \\
\multicolumn{1}{c}{} & \textbf{Sempala} & 29321 & 29684 & 29595 & 29696 & 29658 & 29663 & 29603 & 19357 & 19388 & 19496 & 19867 & 20162 & 20152 & 19737 \\
\multicolumn{1}{c}{} & \textbf{PigSPARQL} & 132130 & 163757 & 181044 & 205225 & 227679 & 255845 & 194280 & 198302 & 252621 & 269446 & 301519 & 324526 & 339143 & 280926 \\
\multicolumn{1}{c}{} & \textbf{SHARD} & 167910 & 189782 & 217487 & 244575 & 275592 & 302084 & 232905 & 167761 & 196993 & 223118 & 252725 & 281614 & 309606 & 238636 \\
\midrule
\multicolumn{1}{c}{\multirow{8}[1]{*}{\begin{sideways}\textbf{SF10000}\end{sideways}}} & \textbf{S2RDF ExtVP} & 12543 & 12252 & 15062 & 15003 & 15478 & 16124 & 14410 & 41188 & 13276 & 14182 & 15261 & 16313 & 13922 & 19024 \\
\multicolumn{1}{c}{} & \textbf{S2RDF VP} & 46996 & 68429 & 69654 & 71388 & 71786 & 73138 & 66899 & 91616 & 118843 & 117261 & 121149 & 124597 & 122869 & 116056 \\
\multicolumn{1}{c}{} & \textbf{H2RDF+} & 76284 & 105794 & 131672 & 164583 & 188800 & 227637 & 149128 & 77567 & 108780 & 139282 & 161913 & 187288 & 216887 & 148620 \\
\multicolumn{1}{c}{} & \textbf{Sempala} & 128486 & 131304 & 152730 & 152169 & 153360 & 154272 & 145387 & 61843 & 63501 & 64487 & 76717 & 97933 & 96590 & 76845 \\
\multicolumn{1}{c}{} & \textbf{PigSPARQL} & 209594 & 270757 & 293241 & 321021 & 348274 & 364243 & 301188 & 258307 & 313681 & 340580 & 365995 & 396331 & 415046 & 348323 \\
\multicolumn{1}{c}{} & \textbf{SHARD} & 792204 & 925542 & 1064010 & 1195541 & 1354956 & 1487522 & 1136629 & 837829 & 992373 & 1131621 & 1278385 & 1414462 & 1622432 & 1212850 \\
\multicolumn{1}{c}{} & \textbf{Virtuoso (cold)} & 46998 & 77903 & 74664 & 82471 & 109177 & 86329 & 79590 & 74014 & 167892 & 78311 & 81350 & 159688 & 95034 & 109382 \\
\multicolumn{1}{c}{} & \textbf{Virtuoso (AM)} & 10529 & 16796 & 13159 & 17320 & 17712 & 18243 & 15627 & 9470  & 19314 & 10775 & 10870 & 18808 & 15431 & 14111 \\
\\
\end{tabularx}%

%% file: experiments/plots/Inc-2.tex
\begin{tabularx}{\textwidth}{rrrrrrrr|r|rrrrrr}
\toprule
\begin{sideways}\textbf{}\end{sideways} & \textbf{Query} & \textbf{IL-3-5} & \textbf{IL-3-6} & \textbf{IL-3-7} & \textbf{IL-3-8} & \textbf{IL-3-9} & \textbf{IL-3-10} & \textbf{AM-IL-3} & \textbf{AM-5} & \textbf{AM-6} & \textbf{AM-7} & \textbf{AM-8} & \textbf{AM-9} & \textbf{AM-10} \\
\midrule
\multicolumn{1}{c}{\multirow{6}[1]{*}{\begin{sideways}\textbf{SF10}\end{sideways}}} & \textbf{S2RDF ExtVP} & 382   & 783   & 732   & 10698 & 1118  & 1241  & 2492  & 381   & 492   & 503   & 3885  & 705   & 868 \\
\multicolumn{1}{c}{} & \textbf{S2RDF VP} & 467   & 1940  & 988   & 11147 & 1254  & 1396  & 2865  & 481   & 1024  & 746   & 4161  & 902   & 998 \\
\multicolumn{1}{c}{} & \textbf{H2RDF+} & 3576  & 6268  & 25927 & 19722 & F     & F     & 13873 & 1596  & 3255  & 9822  & 7736  & N/A   & N/A \\
\multicolumn{1}{c}{} & \textbf{Sempala} & 2624  & 3155  & 1882  & 12548 & 3454  & 3620  & 4547  & 1600  & 1757  & 1373  & 4973  & 1965  & 2085 \\
\multicolumn{1}{c}{} & \textbf{PigSPARQL} & 105205 & 132850 & 145125 & 161200 & 161574 & 179641 & 147599 & 117639 & 147769 & 167219 & 190924 & 208599 & 231707 \\
\multicolumn{1}{c}{} & \textbf{SHARD} & 118258 & 143931 & 158115 & 251074 & 255772 & 273182 & 200055 & 115015 & 136517 & 152063 & 197771 & 210080 & 230115 \\
\midrule
\multicolumn{1}{c}{\multirow{6}[2]{*}{\begin{sideways}\textbf{SF100}\end{sideways}}} & \textbf{S2RDF ExtVP} & 855   & 2766  & 1978  & 36443 & 3608  & 3244  & 8149  & 794   & 1322  & 1114  & 12691 & 1798  & 1863 \\
\multicolumn{1}{c}{} & \textbf{S2RDF VP} & 1365  & 6412  & 2661  & 40297 & 4216  & 4245  & 9866  & 1253  & 3082  & 1896  & 14503 & 2567  & 2644 \\
\multicolumn{1}{c}{} & \textbf{H2RDF+} & 40956 & 70453 & 92907 & F     & F     & F     & 68105 & 29306 & 43815 & 58178 & N/A   & N/A   & N/A \\
\multicolumn{1}{c}{} & \textbf{Sempala} & 19214 & 26118 & 11818 & 111690 & 26654 & 28562 & 37343 & 8340  & 10709 & 5984  & 39353 & 11074 & 11777 \\
\multicolumn{1}{c}{} & \textbf{PigSPARQL} & 164752 & 218817 & 236235 & 345529 & 294268 & 309749 & 261558 & 153507 & 194176 & 215020 & 268601 & 267920 & 287851 \\
\multicolumn{1}{c}{} & \textbf{SHARD} & 205248 & 312420 & 325451 & 1048829 & 884637 & 910991 & 614596 & 146973 & 193078 & 209442 & 464334 & 420804 & 444361 \\
\midrule
\multicolumn{1}{c}{\multirow{6}[2]{*}{\begin{sideways}\textbf{SF1000}\end{sideways}}} & \textbf{S2RDF ExtVP} & 4474  & 12188 & 8552  & 178514 & 13411 & 13405 & 38424 & 3714  & 5267  & 4166  & 60886 & 5993  & 6154 \\
\multicolumn{1}{c}{} & \textbf{S2RDF VP} & 7616  & 41907 & 13993 & 196114 & 22466 & 21487 & 50597 & 6713  & 19088 & 10020 & 70925 & 13364 & 12867 \\
\multicolumn{1}{c}{} & \textbf{H2RDF+} & 121396 & 183752 & 225669 & F     & F     & F     & 176939 & 57279 & 94151 & 118100 & N/A   & N/A   & N/A \\
\multicolumn{1}{c}{} & \textbf{Sempala} & 155298 & 194758 & 93424 & 878232 & 217636 & 231430 & 295130 & 67992 & 81277 & 47505 & 309265 & 89152 & 93748 \\
\multicolumn{1}{c}{} & \textbf{PigSPARQL} & 362172 & 571965 & 622899 & 1924061 & 1653627 & 1777284 & 1152001 & 230868 & 329447 & 357796 & 810268 & 735277 & 790757 \\
\multicolumn{1}{c}{} & \textbf{SHARD} & 1323657 & 2423349 & F     & F     & F     & F     & 1873503 & 553109 & 936708 & N/A   & N/A   & N/A   & N/A \\
\midrule
\multicolumn{1}{c}{\multirow{8}[1]{*}{\begin{sideways}\textbf{SF10000}\end{sideways}}} & \textbf{S2RDF ExtVP} & 29590 & 87525 & 102971 & 2068100 & 158595 & 141940 & 431454 & 27774 & 37684 & 44072 & 699454 & 63462 & 57329 \\
\multicolumn{1}{c}{} & \textbf{S2RDF VP} & 73176 & 316955 & 161557 & 2144949 & 253580 & 247050 & 532878 & 70596 & 168076 & 116157 & 779162 & 149987 & 147686 \\
\multicolumn{1}{c}{} & \textbf{H2RDF+} & 240339 & 451390 & F     & F     & F     & F     & 345865 & 131397 & 221988 & N/A   & N/A   & N/A   & N/A \\
\multicolumn{1}{c}{} & \textbf{Sempala} & 493016 & 595152 & 365868 & 5649620 & 2026680 & 2462137 & 1932079 & 227782 & 263319 & 194362 & 1959502 & 759324 & 904333 \\
\multicolumn{1}{c}{} & \textbf{PigSPARQL} & 1847039 & 3353907 & 4876005 & 40140420 & 37353210 & 37514308 & 20847481 & 771646 & 1312782 & 1836609 & 13609145 & 12699271 & 12764532 \\
\multicolumn{1}{c}{} & \textbf{SHARD} & 11995677 & 23164293 & F     & F     & F     & F     & 17579985 & 4541903 & 8360736 & N/A   & N/A   & N/A   & N/A \\
\multicolumn{1}{c}{} & \textbf{Virtuoso (cold)} & F     & F     & F     & F     & F     & F     & N/A   & N/A   & N/A   & N/A   & N/A   & N/A   & N/A \\
\multicolumn{1}{c}{} & \textbf{Virtuoso (AM)} & F     & F     & F     & F     & F     & F     & N/A   & N/A   & N/A   & N/A   & N/A   & N/A   & N/A \\
\bottomrule
\end{tabularx}%

%% file: sections/conclusion.tex
In this paper, we present S2RDF, a distributed Hadoop-based SPARQL query processor for large-scale RDF data implemented on top of Spark. It comes with a novel relational schema for RDF called ExtVP (\underline{Ext}ended \underline{V}ertical \underline{P}artitioning) and uses the SQL interface of Spark for query execution by compiling SPARQL to SQL.
ExtVP is an extension to the Vertical Partitioning (VP) schema~\cite{abadi_vp_2007} and is inspired by semi-join reductions similar to Join Indices~\cite{valduriez_JoinIndices_1987}. We precompute the reductions of tables in VP for possible join correlations that can occur between triple patterns in a SPARQL query.
In this manner, S2RDF can avoid dangling tuples in the join input which significantly reduces the query input size and thus execution runtime. This technique is applicable to any kind of query pattern regardless of its shape which we demonstrate in our comprehensive evaluation.

S2RDF outperforms state of the art centralized and distributed SPARQL query processors by an order of magnitude on average for all query shapes while achieving sub-second runtimes for majority of benchmark queries on a billion triples dataset. In contrast to most existing distributed RDF stores, the performance of S2RDF using ExtVP does not depend on query diameter but is able to achieve efficient runtimes on large diameter queries as well.
To reduce the size overhead of ExtVP, one can also specify a threshold for selectivity factor $SF$ of ExtVP tables. As we demonstrate in our evaluation, a threshold of 0.25 already achieves $95\%$ of best possible runtime improvements on average while using only $25\%$ of table tuples and storage size.

For future work, we investigate further techniques to reduce the size overhead of ExtVP using a more compact bit vector representation. Furthermore, we are looking at a unification strategy to reduce the number of tables in ExtVP while at the same time being able to consider the intersections of all correlations for a triple pattern in a query which can further improve input selectivity for query execution.

%% file: sections/appendix_watdiv_basic.tex
This is the standard predefined use case in WatDiv covering the full spectrum of SPARQL BGP query shapes.
Terms enclosed within \% are placeholders that get instantiated dynamically by the WatDiv query generator based on the \verb|#mapping| command.

\subsection{Linear Queries}

\begin{itemize}
\item [\textbf{L1:}]
\begin{verbatim}
#mapping v1 wsdbm:Website uniform
SELECT ?v0 ?v2 ?v3 WHERE {
  ?v0  wsdbm:subscribes  %v1% .
  ?v2  sorg:caption	     ?v3 .
  ?v0  wsdbm:likes	     ?v2 .
}
\end{verbatim}

\item [\textbf{L2:}]
\begin{verbatim}
#mapping v0 wsdbm:City uniform
SELECT ?v1 ?v2 WHERE {
  %v0%  gn:parentCountry  ?v1 .
  ?v2   wsdbm:likes       wsdbm:Product0 .
  ?v2   sorg:nationality  ?v1 .
}
\end{verbatim}

\item [\textbf{L3:}]
\begin{verbatim}
#mapping v2 wsdbm:Website uniform
SELECT ?v0 ?v1 WHERE {
  ?v0  wsdbm:likes       ?v1 .
  ?v0  wsdbm:subscribes  %v2% .
}
\end{verbatim}

\item [\textbf{L4:}]
\begin{verbatim}
#mapping v1 wsdbm:Topic uniform
SELECT ?v0 ?v2 WHERE {
  ?v0  og:tag        %v1% .
  ?v0  sorg:caption  ?v2 .
}
\end{verbatim}

\item [\textbf{L5:}]
\begin{verbatim}
#mapping v2 wsdbm:City uniform
SELECT ?v0 ?v1 ?v3 WHERE {
  ?v0   sorg:jobTitle     ?v1 .
  %v2%  gn:parentCountry  ?v3 .
  ?v0   sorg:nationality  ?v3 .
}
\end{verbatim}
\end{itemize}

\subsection{Star Queries}

\begin{itemize}
\item [\textbf{S1:}]
\begin{verbatim}
#mapping v2 wsdbm:Retailer uniform
SELECT ?v0 ?v1 ?v3 ?v4 ?v5
       ?v6 ?v7 ?v8 ?v9 WHERE {
  ?v0   gr:includes            ?v1 .
  %v2%  gr:offers              ?v0 .
  ?v0   gr:price               ?v3 .
  ?v0   gr:serialNumber        ?v4 .
  ?v0   gr:validFrom           ?v5 .
  ?v0   gr:validThrough        ?v6 .
  ?v0   sorg:eligibleQuantity  ?v7 .
  ?v0   sorg:eligibleRegion    ?v8 .
  ?v0   sorg:priceValidUntil   ?v9 .
}
\end{verbatim}

\item [\textbf{S2:}]
\begin{verbatim}
#mapping v2 wsdbm:Country uniform
SELECT ?v0 ?v1 ?v3 WHERE {
  ?v0  dc:Location       ?v1 .
  ?v0  sorg:nationality  %v2% .
  ?v0  wsdbm:gender      ?v3 .
  ?v0  rdf:type          wsdbm:Role2 .
}
\end{verbatim}

\item [\textbf{S3:}]
\begin{verbatim}
#mapping v1 wsdbm:ProductCategory uniform
SELECT ?v0 ?v2 ?v3 ?v4 WHERE {
  ?v0  rdf:type        %v1% .
  ?v0  sorg:caption    ?v2 .
  ?v0  wsdbm:hasGenre  ?v3 .
  ?v0  sorg:publisher  ?v4 .
}
\end{verbatim}

\item [\textbf{S4:}]
\begin{verbatim}
#mapping v1 wsdbm:AgeGroup uniform
SELECT ?v0 ?v2 ?v3 WHERE {
  ?v0  foaf:age          %v1% .
  ?v0  foaf:familyName   ?v2 .
  ?v3  mo:artist         ?v0 .
  ?v0  sorg:nationality  wsdbm:Country1 .
}
\end{verbatim}

\item [\textbf{S5:}]
\begin{verbatim}
#mapping v1 wsdbm:ProductCategory uniform
SELECT ?v0 ?v2 ?v3 WHERE {
  ?v0  rdf:type          %v1% .
  ?v0  sorg:description  ?v2 .
  ?v0  sorg:keywords     ?v3 .
  ?v0  sorg:language     wsdbm:Language0 .
}
\end{verbatim}

\item [\textbf{S6:}]
\begin{verbatim}
#mapping v3 wsdbm:SubGenre uniform
SELECT ?v0 ?v1 ?v2 WHERE {
  ?v0  mo:conductor    ?v1 .
  ?v0  rdf:type        ?v2 .
  ?v0  wsdbm:hasGenre  %v3% .
}
\end{verbatim}

\item [\textbf{S7:}]
\begin{verbatim}
#mapping v3 wsdbm:User uniform
SELECT ?v0 ?v1 ?v2 WHERE {
  ?v0   rdf:type     ?v1 .
  ?v0   sorg:text    ?v2 .
  %v3%  wsdbm:likes  ?v0 .
}
\end{verbatim}
\end{itemize}

\subsection{Snowflake Queries}

\begin{itemize}
\item [\textbf{F1:}]
\begin{verbatim}
#mapping v1 wsdbm:Topic uniform
SELECT ?v0 ?v2 ?v3 ?v4 ?v5 WHERE {
  ?v0  og:tag          %v1% .
  ?v0  rdf:type        ?v2 .
  ?v3  sorg:trailer    ?v4 .
  ?v3  sorg:keywords   ?v5 .
  ?v3  wsdbm:hasGenre  ?v0 .
  ?v3  rdf:type        wsdbm:ProductCategory2 .
}
\end{verbatim}

\item [\textbf{F2:}]
\begin{verbatim}
#mapping v8 wsdbm:SubGenre uniform
SELECT ?v0 ?v1 ?v2 ?v4 ?v5 ?v6 ?v7 WHERE {
  ?v0  foaf:homepage     ?v1 .
  ?v0  og:title          ?v2 .
  ?v0  rdf:type          ?v3 .
  ?v0  sorg:caption      ?v4 .
  ?v0  sorg:description  ?v5 .
  ?v1  sorg:url          ?v6 .
  ?v1  wsdbm:hits        ?v7 .
  ?v0  wsdbm:hasGenre    %v8% .
}
\end{verbatim}

\newpage
\item [\textbf{F3:}]
\begin{verbatim}
#mapping v3 wsdbm:SubGenre uniform
SELECT ?v0 ?v1 ?v2 ?v4 ?v5 ?v6 WHERE {
  ?v0  sorg:contentRating   ?v1 .
  ?v0  sorg:contentSize     ?v2 .
  ?v0  wsdbm:hasGenre       %v3% .
  ?v4  wsdbm:makesPurchase  ?v5 .
  ?v5  wsdbm:purchaseDate   ?v6 .
  ?v5  wsdbm:purchaseFor    ?v0 .
}
\end{verbatim}

\item [\textbf{F4:}]
\begin{verbatim}
#mapping v3 wsdbm:Topic uniform
SELECT ?v0 ?v1 ?v2 ?v4 ?v5 ?v6 ?v7 ?v8 WHERE {
  ?v0  foaf:homepage     ?v1 .
  ?v2  gr:includes       ?v0 .
  ?v0  og:tag            %v3% .
  ?v0  sorg:description  ?v4 .
  ?v0  sorg:contentSize  ?v8 .
  ?v1  sorg:url          ?v5 .
  ?v1  wsdbm:hits        ?v6 .
  ?v1  sorg:language     wsdbm:Language0 .
  ?v7  wsdbm:likes       ?v0 .
}
\end{verbatim}

\item [\textbf{F5:}]
\begin{verbatim}
#mapping v2 wsdbm:Retailer uniform
SELECT ?v0 ?v1 ?v3 ?v4 ?v5 ?v6 WHERE {
  ?v0   gr:includes      ?v1 .
  %v2%  gr:offers        ?v0 .
  ?v0   gr:price         ?v3 .
  ?v0   gr:validThrough  ?v4 .
  ?v1   og:title         ?v5 .
  ?v1   rdf:type         ?v6 .
}
\end{verbatim}
\end{itemize}

\newpage
\subsection{Complex Queries}

\begin{itemize}
\item [\textbf{C1:}]
\begin{verbatim}
SELECT ?v0 ?v4 ?v6 ?v7 WHERE {
  ?v0  sorg:caption        ?v1 .
  ?v0  sorg:text           ?v2 .
  ?v0  sorg:contentRating  ?v3 .
  ?v0  rev:hasReview       ?v4 .
  ?v4  rev:title           ?v5 .
  ?v4  rev:reviewer        ?v6 .
  ?v7  sorg:actor          ?v6 .
  ?v7  sorg:language       ?v8 .
}
\end{verbatim}

\item [\textbf{C2:}]
\begin{verbatim}
SELECT ?v0 ?v3 ?v4 ?v8 WHERE {
  ?v0  sorg:legalName       ?v1 .
  ?v0  gr:offers            ?v2 .
  ?v2  sorg:eligibleRegion  wsdbm:Country5 .
  ?v2  gr:includes          ?v3 .
  ?v4  sorg:jobTitle        ?v5 .
  ?v4  foaf:homepage        ?v6 .
  ?v4  wsdbm:makesPurchase  ?v7 .
  ?v7  wsdbm:purchaseFor    ?v3 .
  ?v3  rev:hasReview        ?v8 .
  ?v8  rev:totalVotes       ?v9 .
}
\end{verbatim}

\item [\textbf{C3:}]
\begin{verbatim}
SELECT ?v0 WHERE {
  ?v0  wsdbm:likes     ?v1 .
  ?v0  wsdbm:friendOf  ?v2 .
  ?v0  dc:Location     ?v3 .
  ?v0  foaf:age        ?v4 .
  ?v0  wsdbm:gender    ?v5 .
  ?v0  foaf:givenName  ?v6 .
}
\end{verbatim}
\end{itemize}

%% file: sections/appendix_watdiv_selectivity.tex
We defined this workload for WatDiv to test the effect of varying $SF$ values of ExtVP tables on query performance.

\subsection{Varying OS Selectivity}

\begin{itemize}
\item [\textbf{1-1:}]
\begin{verbatim}
# ?v0 -- friendOf --> ?v1 -- email --> ?v2
# |VP_fiendOf| = 0.41 * |G|
# SF(ExtVP_OS_fiendOf|email) = 0.90
# SF(ExtVP_SO_email|fiendOf) = 1
SELECT ?v0 ?v1 ?v2 WHERE {
  ?v0  wsdbm:friendOf  ?v1 .
  ?v1  sorg:email      ?v2 .
}
\end{verbatim}

\item [\textbf{1-2:}]
\begin{verbatim}
# ?v0 -- friendOf --> ?v1 -- age --> ?v2
# |VP_fiendOf| = 0.41 * |G|
# SF(ExtVP_OS_fiendOf|age) = 0.50
# SF(ExtVP_SO_age|fiendOf) = 1
SELECT ?v0 ?v1 ?v2 WHERE {
  ?v0  wsdbm:friendOf  ?v1 .
  ?v1  foaf:age        ?v2 .
}
\end{verbatim}

\item [\textbf{1-3:}]
\begin{verbatim}
# ?v0 -- friendOf --> ?v1 -- jobTitle --> ?v2
# |VP_fiendOf| = 0.41 * |G|
# SF(ExtVP_OS_fiendOf|jobTitle) = 0.05
# SF(ExtVP_SO_jobTitle|fiendOf) = 1
SELECT ?v0 ?v1 ?v2 WHERE {
  ?v0  wsdbm:friendOf  ?v1 .
  ?v1  sorg:jobTitle   ?v2 .
}
\end{verbatim}

\item [\textbf{2-1:}]
\begin{verbatim}
# ?v0 -- reviewer --> ?v1 -- email --> ?v2
# |VP_reviewer| = 0.01 * |G|
# SF(ExtVP_OS_reviewer|email) = 0.90
# SF(ExtVP_SO_email|reviewer) = 0.31
SELECT ?v0 ?v1 ?v2 WHERE {
  ?v0  rev:reviewer  ?v1 .
  ?v1  sorg:email    ?v2 .
}
\end{verbatim}

\item [\textbf{2-2:}]
\begin{verbatim}
# ?v0 -- reviewer --> ?v1 -- age --> ?v2
# |VP_reviewer| = 0.01 * |G|
# SF(ExtVP_OS_reviewer|age) = 0.50
# SF(ExtVP_SO_age|reviewer) = 0.31
SELECT ?v0 ?v1 ?v2 WHERE {
  ?v0  rev:reviewer  ?v1 .
  ?v1  foaf:age      ?v2 .
}
\end{verbatim}

\item [\textbf{2-3:}]
\begin{verbatim}
# ?v0 -- reviewer --> ?v1 -- jobTitle --> ?v2
# |VP_reviewer| = 0.01 * |G|
# SF(ExtVP_OS_reviewer|jobTitle) = 0.05
# SF(ExtVP_SO_jobTitle|reviewer) = 0.31
SELECT ?v0 ?v1 ?v2 WHERE {
  ?v0  rev:reviewer   ?v1 .
  ?v1  sorg:jobTitle  ?v2 .
}
\end{verbatim}
\end{itemize}

\newpage
\subsection{Varying SO Selectivity}

\begin{itemize}
\item [\textbf{3-1:}]
\begin{verbatim}
# ?v0 -- follows --> ?v1 -- friendOf --> ?v2
# |VP_fiendOf| = 0.41 * |G|
# SF(ExtVP_OS_follows|fiendOf) = 0.40
# SF(ExtVP_SO_fiendOf|follows) = 0.90
SELECT ?v0 ?v1 ?v2 WHERE {
  ?v0  wsdbm:follows   ?v1 .
  ?v1  wsdbm:friendOf  ?v2 .
}
\end{verbatim}

\item [\textbf{3-2:}]
\begin{verbatim}
# ?v0 -- reviewer --> ?v1 -- friendOf --> ?v2
# |VP_fiendOf| = 0.41 * |G|
# SF(ExtVP_OS_reviewer|fiendOf) = 0.40
# SF(ExtVP_SO_fiendOf|reviewer) = 0.31
SELECT ?v0 ?v1 ?v2 WHERE {
  ?v0  rev:reviewer    ?v1 .
  ?v1  wsdbm:friendOf  ?v2 .
}
\end{verbatim}

\item [\textbf{3-3:}]
\begin{verbatim}
# ?v0 -- author --> ?v1 -- friendOf --> ?v2
# |VP_fiendOf| = 0.41 * |G|
# SF(ExtVP_OS_author|fiendOf) = 0.40
# SF(ExtVP_SO_fiendOf|author) = 0.04
SELECT ?v0 ?v1 ?v2 WHERE {
  ?v0  sorg:author     ?v1 .
  ?v1  wsdbm:friendOf  ?v2 .
}
\end{verbatim}

\item [\textbf{4-1:}]
\begin{verbatim}
# ?v0 -- follows --> ?v1 -- likes --> ?v2
# |VP_likes| = 0.01 * |G|
# SF(ExtVP_OS_follows|likes) = 0.24
# SF(ExtVP_SO_likes|follows) = 0.90
SELECT ?v0 ?v1 ?v2 WHERE {
  ?v0  wsdbm:follows  ?v1 .
  ?v1  wsdbm:likes    ?v2 .
}
\end{verbatim}

\item [\textbf{4-2:}]
\begin{verbatim}
# ?v0 -- reviewer --> ?v1 -- likes --> ?v2
# |VP_likes| = 0.01 * |G|
# SF(ExtVP_OS_reviewer|likes) = 0.24
# SF(ExtVP_SO_likes|reviewer) = 0.31
SELECT ?v0 ?v1 ?v2 WHERE {
  ?v0  wsdbm:reviewer  ?v1 .
  ?v1  wsdbm:likes     ?v2 .
}
\end{verbatim}

\item [\textbf{4-3:}]
\begin{verbatim}
# ?v0 -- author --> ?v1 -- likes --> ?v2
# |VP_likes| = 0.01 * |G|
# SF(ExtVP_OS_author|likes) = 0.24
# SF(ExtVP_SO_likes|author) = 0.04
SELECT ?v0 ?v1 ?v2 WHERE {
  ?v0  wsdbm:author  ?v1 .
  ?v1  wsdbm:likes   ?v2 .
}
\end{verbatim}
\end{itemize}

\newpage
\subsection{Varying SS Selectivity}

\begin{itemize}
\item [\textbf{5-1:}]
\begin{verbatim}
# ?v0 <-- friendOf -- ?v1 -- email --> ?v2
# |VP_fiendOf| = 0.41 * |G|
# SF(ExtVP_SS_fiendOf|email) = 0.90
# SF(ExtVP_SS_email|fiendOf) = 0.40
SELECT ?v0 ?v1 ?v2 WHERE {
  ?v0  wsdbm:friendOf  ?v1 .
  ?v0  sorg:email      ?v2 .
}
\end{verbatim}

\item [\textbf{5-2:}]
\begin{verbatim}
# ?v0 <-- friendOf -- ?v1 -- follows --> ?v2
# |VP_fiendOf| = 0.41 * |G|
# SF(ExtVP_SS_fiendOf|follows) = 0.77
# SF(ExtVP_SS_follows|fiendOf) = 0.40
SELECT ?v0 ?v1 ?v2 WHERE {
  ?v0  wsdbm:friendOf  ?v1 .
  ?v0  wsdbm:follows   ?v2 .
}
\end{verbatim}
\end{itemize}

\subsection{High Selectivity Queries}

\begin{itemize}
\item [\textbf{6-1:}]
\begin{verbatim}
# ?v0 -- likes --> ?v1 -- trailer --> ?v2
# |VP_likes| = 0.01 * |G|
# |VP_trailer| = < 0.01 * |G|
# SF(ExtVP_OS_likes|trailer) < 0.01
# SF(ExtVP_SO_trailer|likes) = 0.96
SELECT ?v0 ?v1 ?v2 WHERE {
  ?v0  wsdbm:likes   ?v1 .
  ?v1  sorg:trailer  ?v2 .
}
\end{verbatim}

\item [\textbf{6-2:}]
\begin{verbatim}
# ?v1 <-- email -- ?v0 -- faxNumber --> ?v2
# |VP_email| = 0.01 * |G|
# |VP_faxNumber| = < 0.01 * |G|
# SF(ExtVP_SS_email|faxNumber) < 0.01
# SF(ExtVP_SS_faxNumber|email) = 0.80
SELECT ?v0 ?v1 ?v2 WHERE {
  ?v0  sorg:email      ?v1 .
  ?v0  sorg:faxNumber  ?v2 .
}
\end{verbatim}
\end{itemize}

\newpage
\subsection{OS vs SO Selectivity}

\begin{itemize}
\item [\textbf{7-1:}]
\begin{verbatim}
# ?v0 -- friendOf --> ?v1 -- follows -->
  ?v2 -- homepage --> ?v3
# SF(ExtVP_SO_follows|friendOf) = 1
# SF(ExtVP_OS_follows|homepage) = 0.05
SELECT ?v0 ?v1 ?v2 ?v3 WHERE {
  ?v0  wsdbm:friendOf  ?v1 .
  ?v1  wsdbm:follows   ?v2 .
  ?v2  foaf:homepage   ?v3 .
}
\end{verbatim}

\item [\textbf{7-2:}]
\begin{verbatim}
# ?v0 -- artist --> ?v1 -- friendOf -->
  ?v2 -- follows --> ?v3
# SF(ExtVP_SO_friendOf|artist) = 0.01
# SF(ExtVP_OS_friendOf|follows) = 0.77
SELECT ?v0 ?v1 ?v2 ?v3 WHERE {
  ?v0  mo:artist       ?v1 .
  ?v1  wsdbm:friendOf  ?v2 .
  ?v2  wsdbm:follows   ?v3 .
}
\end{verbatim}
\end{itemize}

\subsection{Empty Result Queries}

\begin{itemize}
\item [\textbf{8-1:}]
\begin{verbatim}
# ?v0 -- friendOf --> ?v1 -- language --> ?v2
# SF(ExtVP_OS_friendOf|language) = 0
# SF(ExtVP_SO_language|friendOf) = 0
SELECT ?v0 ?v1 ?v2 WHERE {
  ?v0  wsdbm:friendOf  ?v1 .
  ?v1  sorg:language   ?v2 .
}
\end{verbatim}

\item [\textbf{8-2:}]
\begin{verbatim}
# ?v0 -- friendOf --> ?v1 -- follows -->
  ?v2 -- language --> ?v3
# SF(ExtVP_OS_follows|language) = 0
# SF(ExtVP_SO_language|follows) = 0
SELECT ?v0 ?v1 ?v2 ?v3 WHERE {
  ?v0  wsdbm:friendOf  ?v1 .
  ?v1  wsdbm:follows   ?v2 .
  ?v2  sorg:language   ?v3 .
}
\end{verbatim}
\end{itemize}

%% file: sections/appendix_watdiv_incremental.tex
We defined this workload for WatDiv to test the performance of S2RDF for linear queries with increasing size (number of triple patterns). The workload contains 3 types of queries (user, retailer and unbound) where we incrementally add triple patterns to the initial query (starting from 5 up to 10 triple patterns). For example, query 1-6 adds another triple pattern to query 1-5 and hence expands the path defined by the query by one.
The workload is now also officially included in WatDiv as an additional use case.

Terms enclosed within \% are placeholders that get instantiated dynamically by the WatDiv query generator based on the \verb|#mapping| command.

\subsection{Incremental User Queries (Type 1)}

\begin{itemize}
\item [\textbf{1-5:}]
\begin{verbatim}
#mapping v0 wsdbm:User uniform
SELECT ?v1 ?v2 ?v3 ?v4 ?v5 WHERE {
  %v0%  wsdbm:follows   ?v1 .
  ?v1   wsdbm:likes     ?v2 .
  ?v2   rev:hasReview   ?v3 .
  ?v3   rev:reviewer    ?v4 .
  ?v4   wsdbm:friendOf  ?v5 .
}
\end{verbatim}

\item [\textbf{1-6:}]
\begin{verbatim}
#mapping v0 wsdbm:User uniform
SELECT ?v1 ?v2 ?v3 ?v4 ?v5 ?v6 WHERE {
  %v0%  wsdbm:follows        ?v1 .
  ?v1   wsdbm:likes          ?v2 .
  ?v2   rev:hasReview        ?v3 .
  ?v3   rev:reviewer         ?v4 .
  ?v4   wsdbm:friendOf       ?v5 .
  ?v5   wsdbm:makesPurchase  ?v6 .
}
\end{verbatim}

\item [\textbf{1-7:}]
\begin{verbatim}
#mapping v0 wsdbm:User uniform
SELECT ?v1 ?v2 ?v3 ?v4 ?v5 ?v6
       ?v7 WHERE {
  %v0%  wsdbm:follows        ?v1 .
  ?v1   wsdbm:likes          ?v2 .
  ?v2   rev:hasReview        ?v3 .
  ?v3   rev:reviewer         ?v4 .
  ?v4   wsdbm:friendOf       ?v5 .
  ?v5   wsdbm:makesPurchase  ?v6 .
  ?v6   wsdbm:purchaseFor    ?v7 .
}
\end{verbatim}

\item [\textbf{1-8:}]
\begin{verbatim}
#mapping v0 wsdbm:User uniform
SELECT ?v1 ?v2 ?v3 ?v4 ?v5 ?v6
       ?v7 ?v8 WHERE {
  %v0%  wsdbm:follows        ?v1 .
  ?v1   wsdbm:likes          ?v2 .
  ?v2   rev:hasReview        ?v3 .
  ?v3   rev:reviewer         ?v4 .
  ?v4   wsdbm:friendOf       ?v5 .
  ?v5   wsdbm:makesPurchase  ?v6 .
  ?v6   wsdbm:purchaseFor    ?v7 .
  ?v7   sorg:author          ?v8 .
}
\end{verbatim}

\newpage
\item [\textbf{1-9:}]
\begin{verbatim}
#mapping v0 wsdbm:User uniform
SELECT ?v1 ?v2 ?v3 ?v4 ?v5 ?v6
       ?v7 ?v8 ?v9 WHERE {
  %v0%  wsdbm:follows        ?v1 .
  ?v1   wsdbm:likes          ?v2 .
  ?v2   rev:hasReview        ?v3 .
  ?v3   rev:reviewer         ?v4 .
  ?v4   wsdbm:friendOf       ?v5 .
  ?v5   wsdbm:makesPurchase  ?v6 .
  ?v6   wsdbm:purchaseFor    ?v7 .
  ?v7   sorg:author          ?v8 .
  ?v8   dc:Location          ?v9 .
}
\end{verbatim}

\item [\textbf{1-10:}]
\begin{verbatim}
#mapping v0 wsdbm:User uniform
SELECT ?v1 ?v2 ?v3 ?v4 ?v5 ?v6
       ?v7 ?v8 ?v9 ?v10 WHERE {
  %v0%  wsdbm:follows        ?v1 .
  ?v1   wsdbm:likes          ?v2 .
  ?v2   rev:hasReview        ?v3 .
  ?v3   rev:reviewer         ?v4 .
  ?v4   wsdbm:friendOf       ?v5 .
  ?v5   wsdbm:makesPurchase  ?v6 .
  ?v6   wsdbm:purchaseFor    ?v7 .
  ?v7   sorg:author          ?v8 .
  ?v8   dc:Location          ?v9 .
  ?v9   gn:parentCountry     ?v10 .
}
\end{verbatim}
\end{itemize}

\subsection{Incremental Retailer Queries (Type 2)}

\begin{itemize}
\item [\textbf{2-5:}]
\begin{verbatim}
#mapping v0 wsdbm:Retailer uniform
SELECT ?v1 ?v2 ?v3 ?v4 ?v5 WHERE {
  %v0%  gr:offers       ?v1 .
  ?v1   gr:includes     ?v2 .
  ?v2   sorg:director   ?v3 .
  ?v3   wsdbm:friendOf  ?v4 .
  ?v4   wsdbm:friendOf  ?v5 .
}
\end{verbatim}

\item [\textbf{2-6:}]
\begin{verbatim}
#mapping v0 wsdbm:Retailer uniform
SELECT ?v1 ?v2 ?v3 ?v4 ?v5 ?v6 WHERE {
  %v0%  gr:offers       ?v1 .
  ?v1   gr:includes     ?v2 .
  ?v2   sorg:director   ?v3 .
  ?v3   wsdbm:friendOf  ?v4 .
  ?v4   wsdbm:friendOf  ?v5 .
  ?v5   wsdbm:likes     ?v6 .
}
\end{verbatim}

\item [\textbf{2-7:}]
\begin{verbatim}
#mapping v0 wsdbm:Retailer uniform
SELECT ?v1 ?v2 ?v3 ?v4 ?v5 ?v6
       ?v7 WHERE {
  %v0%  gr:offers       ?v1 .
  ?v1   gr:includes     ?v2 .
  ?v2   sorg:director   ?v3 .
  ?v3   wsdbm:friendOf  ?v4 .
  ?v4   wsdbm:friendOf  ?v5 .
  ?v5   wsdbm:likes     ?v6 .
  ?v6   sorg:editor     ?v7 .
}
\end{verbatim}

\newpage
\item [\textbf{2-8:}]
\begin{verbatim}
#mapping v0 wsdbm:Retailer uniform
SELECT ?v1 ?v2 ?v3 ?v4 ?v5 ?v6
       ?v7 ?v8 WHERE {
  %v0%  gr:offers            ?v1 .
  ?v1   gr:includes          ?v2 .
  ?v2   sorg:director        ?v3 .
  ?v3   wsdbm:friendOf       ?v4 .
  ?v4   wsdbm:friendOf       ?v5 .
  ?v5   wsdbm:likes          ?v6 .
  ?v6   sorg:editor          ?v7 .
  ?v7   wsdbm:makesPurchase  ?v8 .
}
\end{verbatim}

\item [\textbf{2-9:}]
\begin{verbatim}
#mapping v0 wsdbm:Retailer uniform
SELECT ?v1 ?v2 ?v3 ?v4 ?v5 ?v6
       ?v7 ?v8 ?v9 WHERE {
  %v0%  gr:offers            ?v1 .
  ?v1   gr:includes          ?v2 .
  ?v2   sorg:director        ?v3 .
  ?v3   wsdbm:friendOf       ?v4 .
  ?v4   wsdbm:friendOf       ?v5 .
  ?v5   wsdbm:likes          ?v6 .
  ?v6   sorg:editor          ?v7 .
  ?v7   wsdbm:makesPurchase  ?v8 .
  ?v8   wsdbm:purchaseFor    ?v9 .
}
\end{verbatim}

\item [\textbf{2-10:}]
\begin{verbatim}
#mapping v0 wsdbm:Retailer uniform
SELECT ?v1 ?v2 ?v3 ?v4 ?v5 ?v6
       ?v7 ?v8 ?v9 ?v10 WHERE {
  %v0%  gr:offers            ?v1 .
  ?v1   gr:includes          ?v2 .
  ?v2   sorg:director        ?v3 .
  ?v3   wsdbm:friendOf       ?v4 .
  ?v4   wsdbm:friendOf       ?v5 .
  ?v5   wsdbm:likes          ?v6 .
  ?v6   sorg:editor          ?v7 .
  ?v7   wsdbm:makesPurchase  ?v8 .
  ?v8   wsdbm:purchaseFor    ?v9 .
  ?v9   sorg:caption         ?v10 .
}
\end{verbatim}
\end{itemize}

\subsection{Incremental Unbound Queries (Type 3)}

\begin{itemize}
\item [\textbf{3-5:}]
\begin{verbatim}
SELECT ?v0 ?v1 ?v2 ?v3 ?v4 ?v5 WHERE {
  ?v0  gr:offers       ?v1 .
  ?v1  gr:includes     ?v2 .
  ?v2  rev:hasReview   ?v3 .
  ?v3  rev:reviewer    ?v4 .
  ?v4  wsdbm:friendOf  ?v5 .
}
\end{verbatim}

\item [\textbf{3-6:}]
\begin{verbatim}
SELECT ?v0 ?v1 ?v2 ?v3 ?v4 ?v5 ?v6 WHERE {
  ?v0  gr:offers       ?v1 .
  ?v1  gr:includes     ?v2 .
  ?v2  rev:hasReview   ?v3 .
  ?v3  rev:reviewer    ?v4 .
  ?v4  wsdbm:friendOf  ?v5 .
  ?v5  wsdbm:likes     ?v6 .
}
\end{verbatim}

\newpage
\item [\textbf{3-7:}]
\begin{verbatim}
SELECT ?v0 ?v1 ?v2 ?v3 ?v4 ?v5 ?v6
       ?v7 WHERE {
  ?v0  gr:offers       ?v1 .
  ?v1  gr:includes     ?v2 .
  ?v2  rev:hasReview   ?v3 .
  ?v3  rev:reviewer    ?v4 .
  ?v4  wsdbm:friendOf  ?v5 .
  ?v5  wsdbm:likes     ?v6 .
  ?v6  sorg:author     ?v7 .
}
\end{verbatim}

\item [\textbf{3-8:}]
\begin{verbatim}
SELECT ?v0 ?v1 ?v2 ?v3 ?v4 ?v5 ?v6
       ?v7 ?v8 WHERE {
  ?v0  gr:offers       ?v1 .
  ?v1  gr:includes     ?v2 .
  ?v2  rev:hasReview   ?v3 .
  ?v3  rev:reviewer    ?v4 .
  ?v4  wsdbm:friendOf  ?v5 .
  ?v5  wsdbm:likes     ?v6 .
  ?v6  sorg:author     ?v7 .
  ?v7  wsdbm:follows   ?v8 .
}
\end{verbatim}

\item [\textbf{3-9:}]
\begin{verbatim}
SELECT ?v0 ?v1 ?v2 ?v3 ?v4 ?v5 ?v6
       ?v7 ?v8 ?v9 WHERE {
  ?v0  gr:offers       ?v1 .
  ?v1  gr:includes     ?v2 .
  ?v2  rev:hasReview   ?v3 .
  ?v3  rev:reviewer    ?v4 .
  ?v4  wsdbm:friendOf  ?v5 .
  ?v5  wsdbm:likes     ?v6 .
  ?v6  sorg:author     ?v7 .
  ?v7  wsdbm:follows   ?v8 .
  ?v8  foaf:homepage   ?v9 .
}
\end{verbatim}

\item [\textbf{3-10:}]
\begin{verbatim}
SELECT ?v0 ?v1 ?v2 ?v3 ?v4 ?v5 ?v6
       ?v7 ?v8 ?v9 ?v10 WHERE {
  ?v0  gr:offers       ?v1 .
  ?v1  gr:includes     ?v2 .
  ?v2  rev:hasReview   ?v3 .
  ?v3  rev:reviewer    ?v4 .
  ?v4  wsdbm:friendOf  ?v5 .
  ?v5  wsdbm:likes     ?v6 .
  ?v6  sorg:author     ?v7 .
  ?v7  wsdbm:follows   ?v8 .
  ?v8  foaf:homepage   ?v9 .
  ?v9  sorg:language   ?v10 .
}
\end{verbatim}
\end{itemize}